\documentclass[twocolumn,aps,prx,reprint,superscriptaddress,longbibliography]{revtex4-2}

\usepackage{dcolumn}    
\usepackage{bm}         
\usepackage{xfrac}      
\usepackage{physics}    
\usepackage{amsmath}
\usepackage{amssymb}
\usepackage{graphicx}
\usepackage{color}
\usepackage{hyperref}
\usepackage{xcolor}
\usepackage{lipsum}
\usepackage{tabularx}
\usepackage{booktabs}
\usepackage{multirow}
\usepackage{enumitem}
\usepackage{titlesec}
\usepackage{soul}

\hypersetup{
    colorlinks,
    linkcolor={blue!80!black},
    citecolor={blue!80!black},
    urlcolor={blue!80!black}
}

\usepackage[caption=false]{subfig}

\begin{document}

\title{Strong coupling of a microwave photon to an electron on helium}

\author{G.~Koolstra}
\thanks{These authors contributed equally to this work}
\affiliation{EeroQ Corporation, Chicago, Illinois, 60651, USA}

\author{E.~O.~Glen}
\thanks{These authors contributed equally to this work}
\affiliation{EeroQ Corporation, Chicago, Illinois, 60651, USA}

\author{N.~R.~Beysengulov}
\thanks{These authors contributed equally to this work}
\affiliation{EeroQ Corporation, Chicago, Illinois, 60651, USA}

\author{H.~Byeon}
\affiliation{EeroQ Corporation, Chicago, Illinois, 60651, USA}

\author{K.~E.~Castoria}
\affiliation{EeroQ Corporation, Chicago, Illinois, 60651, USA}

\author{M.~Sammon}
\affiliation{EeroQ Corporation, Chicago, Illinois, 60651, USA}

\author{S.~A.~Lyon}
\affiliation{EeroQ Corporation, Chicago, Illinois, 60651, USA}

\author{D.~G.~Rees}
\affiliation{EeroQ Corporation, Chicago, Illinois, 60651, USA}

\author{J.~Pollanen}
\affiliation{EeroQ Corporation, Chicago, Illinois, 60651, USA}

\date{\today}

\begin{abstract}
Electrons bound to the surface of superfluid helium have been proposed for scalable charge and spin-based quantum computing. However single electron quantum measurement in this system has remained elusive. Here we use a hybrid circuit quantum electrodynamic (cQED) device that comprises a quantum dot and a high-impedance superconducting resonator to demonstrate, for the first time, strong coupling between the resonator microwave field and the motional quantum state of the electron. We find a coupling strength between the electron motion and a resonator photon of $g/2\pi=118$~MHz, exceeding both the electron motional state decoherence and the resonator loss. These experiments open new avenues for investigating light-matter interaction at the single electron level, and are a key step towards measurement and control of electrons on helium-based spin qubits.     
\end{abstract}

\maketitle

Electrons trapped above the surface of superfluid helium are a compelling system for quantum information science and computing~\cite{Dykman1999, Dykman2003,Schuster2010,Lyon2006,Kawakami2025}. Electrons confined in nanofabricated traps at the helium surface exhibit quantized motional states~\cite{Schuster2010}, and share many of the same qualities found in quantum dot-based devices in semiconductors, while offering the added advantage of being hosted on a pristine substrate free of local charge traps or nuclear spins. Electrons on helium can be integrated with monolithic device architectures to demonstrate the high-fidelity shuttling~\cite{Bradbury2011} and precision single electron control~\cite{Castoria2025, Koolstra2019} needed for scalable devices. Additionally, the electron spin states on helium are predicted to have extremely low spin-orbit interaction and long spin coherence times exceeding 10 seconds~\cite{Lyon2006}, making them an attractive candidate for use as spin qubits. 

One particularly promising avenue for spin readout of electrons on helium involves coupling the electron motional state to a superconducting microwave resonator using a circuit quantum electrodynamic (cQED) scheme, while inducing spin-charge hybridization with a local magnetic field gradient~\cite{Schuster2010}. However, successful demonstrations in other systems have typically leveraged strong coupling to the motional state, in which the electron–resonator interaction rate ($g$) surpasses both the electron decoherence ($\Gamma_{2}$) and resonator linewidth ($\kappa$). In semiconductor systems, for example, reaching the strong electron motional coupling regime has provided a powerful route for subsequent spin-readout demonstrations~\cite{Samkharadze2018, Mi2018}.

While the motional states of an electron on helium have been coupled to superconducting microwave cavities~\cite{Koolstra2019, Castoria2025}, strong charge-photon coupling has remained elusive due to the weak interaction between the resonator field and the quantized motion of the confined electron~\cite{Koolstra2019}. Here we report on experiments that demonstrate strong coupling of the electron motional state in a quantum dot to an on-chip microwave resonator embedded in a microfluidic device architecture. We measure an electron-resonator coupling rate of $g/2\pi=118$~MHz, which is larger than the resonator linewidth $\kappa/2\pi=23$~MHz and electron decoherence rate $\Gamma_{2}/2\pi=61$~MHz. We demonstrate precise control over individual electrons as well as their motional states within the quantum dot, in excellent agreement with finite element modeling (FEM) of the device. Furthermore, we show that electrostatic control of the confining potential of the dot allows us to tune the electron charge state frequency and its coherence. These results pave a path towards electron on helium-based quantum computing architectures as well as new opportunities for investigating the strong interaction of light and matter at the level of a single electron. 

\begin{figure*} 
\centering
	\includegraphics[width=0.9\linewidth]{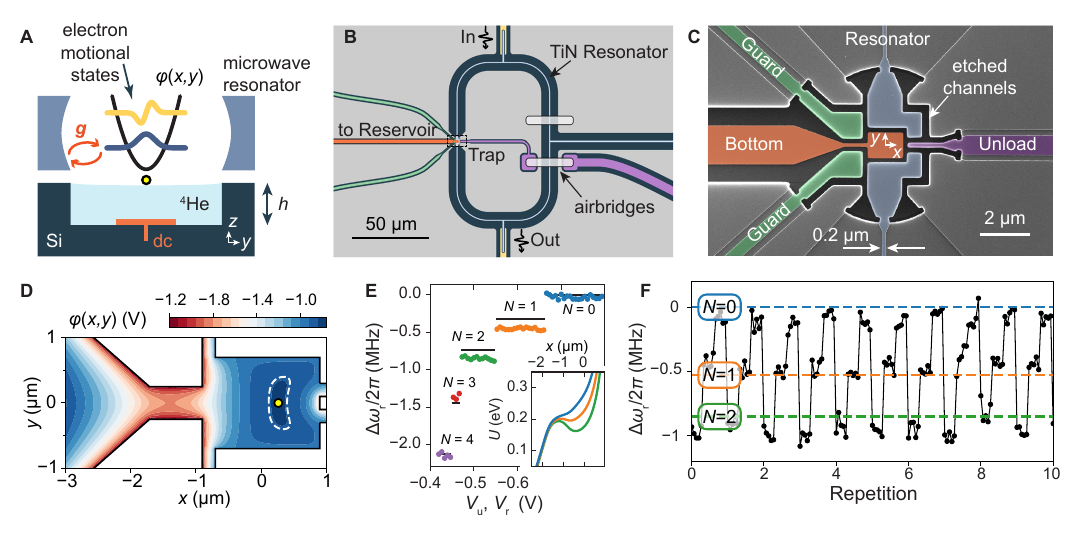}
	\caption{\textbf{Hybrid electron on helium-cQED device} (\textbf{A}) Schematic of electron motional states coupled to microwave photons in a resonator. Electron is laterally confined above helium surface by trapping potential $\varphi (x,y)$ generated from DC voltages on electrodes. (\textbf{B}) TiN resonator layout with integrated electron dot. Air bridges connect the ground plane and facilitate electrode connections. (\textbf{C}) False-colored scanning electron microscope image of electron trapping region with control electrodes highlighted in color. (\textbf{D}) Typical map of $\varphi (x,y)$ in the dot, calculated using FEM. The voltages are tuned to confine a single electron, which promotes the quantization axis to be aligned with the $y$-axis, maximizing its coupling to microwave photons. (\textbf{E}) Steps in the resonator frequency shifts $\Delta \omega_{r}$ measure the number of electrons the dot (black lines are modeled values from FEM). Inset shows modeled electrostatic potential energy $U$ along the channel for $V_{r} = V_{u} = -0.7$~V (blue, $N = 0$), $-0.6$~V (orange, $N = 1$) and $-0.5$~V (green, $N = 2$). (\textbf{F}) Consistent and repeated loading of $N=2, 1$ and 0 electrons into and out of dot. Dashed lines indicate $\Delta \omega_{r}$ for $N=2, 1, 0$ electrons from (E).}
\label{fig:fig1}
\end{figure*}

An electron trapped above the surface of superfluid helium is bound to its induced image charge in the helium while experiencing a $\sim$1~eV barrier that prevents it from penetrating into the liquid. This provides out of plane confinement, but allows for lateral control of the electron along the helium surface using electrostatic voltages in silicon-based microfluidic devices~\cite{glasson2001observation}. The device used for these experiments consists of a single electron on helium, held in a trapping potential that creates a quantum dot, and coupled to an on-chip high-impedance coplanar microwave resonator (Fig.~\ref{fig:fig1}\,A-C). Electrons are transported into the dot from an on-chip reservoir (see Materials and Methods). In this work we design a more compact quantum dot with dimensions of $1.4 \times 1.4\ \mu\text{m}^2$, and implement a high-kinetic inductance resonator with impedance of $Z=3.8$~k$\Omega$ to boost the coupling energy by a factor of 20 compared to previous realizations of electron on helium quantum dots~\cite{Koolstra2019}, resulting in a designed electron-photon interaction rate of $g/2\pi = 110$~MHz (Supplementary Text, S\ref{supp:g}).
The quantized motional frequency of the electron $\omega_{e}/2\pi$ is tuned by the electrostatic potential in the dot, which is defined by the device geometry and controlled by a combination of voltages $V_{b}, V_{r}, V_{u}, V_{g},$ applied to the Bottom, Resonator, Unload and Guard electrodes, respectively (Fig.~\ref{fig:fig1}\,C,D). The Bottom electrode is below the helium level and is used to transport electrons from the reservoir to the dot (see Materials and Methods). On the top plane of the device, the Guard electrodes provide a potential barrier between the electrons in the dot and those in the reservoir (Fig.~\ref{fig:fig1}\,D). The resonator consists of two titanium nitride (TiN) arms attached to a common bias line for supplying a DC voltage, as described in Ref.~\cite{Koolstra2025,Castoria2025}. 
The differential mode of the resonator has a frequency of $\omega_{r}/2\pi=7.162$ GHz and linewidth $\kappa/2\pi=23$ MHz, which we determine from measurements of the resonator transmission $S_{21}$ (see Materials and Methods). The microwave field of the differential mode couples to the electron motion in the dot allowing the resonator to be used as a sensitive electron detector~\cite{Koolstra2019,Zhou2022,Castoria2025}.

We first demonstrate deterministic control of the number of electrons in the dot, enabled by the smooth, defect-free helium surface~\cite{Bradbury2011}. The dot is initially loaded with several electrons from the reservoir (see Materials and Methods). Electrons are unloaded one at a time as the trapping potential depth is reduced by setting $V_{r}$ and $V_{u}$ progressively more negative. Between each unloading event, we return the bias to a probe condition ($V_{r}=0.1$~V, $V_{u} = -0.3$~V) and measure the resonator frequency shift $\Delta \omega_{r}/2\pi$ at the same electrostatic curvature (Fig.~\ref{fig:fig1}\,E). We infer the number of electrons by measuring the resonator frequency shift $\Delta \omega_{r}/2\pi$ after each unloading cycle (Supplementary Text, S\ref{supp:electron_cluster_properties}). For the most negative $V_{u}$ and $V_{r}$, the dot is empty and the resonator returns to its bare resonance ($\Delta \omega_{r}/2\pi=0$). To showcase the precise control we have over the electron number, we repeatedly isolate $N=2$ electrons from the reservoir and subsequently unload to $N= 1$ and 0 electrons (Fig.~\ref{fig:fig1}\,E). Additionally, electrons can be trapped indefinitely because the depth of the electrostatic trap of the quantum dot ($\sim10^3$~GHz) exceeds both the electron motional state frequency ($\simeq$4-8~GHz) and the thermal energy ($\simeq$0.2~GHz) by several orders of magnitude, such that quantum tunneling does not play a role (Supplementary Text, figure~\ref{fig:supp:energy_scales}). 

\begin{figure*} 
\centering
	\includegraphics[width=0.7\linewidth]{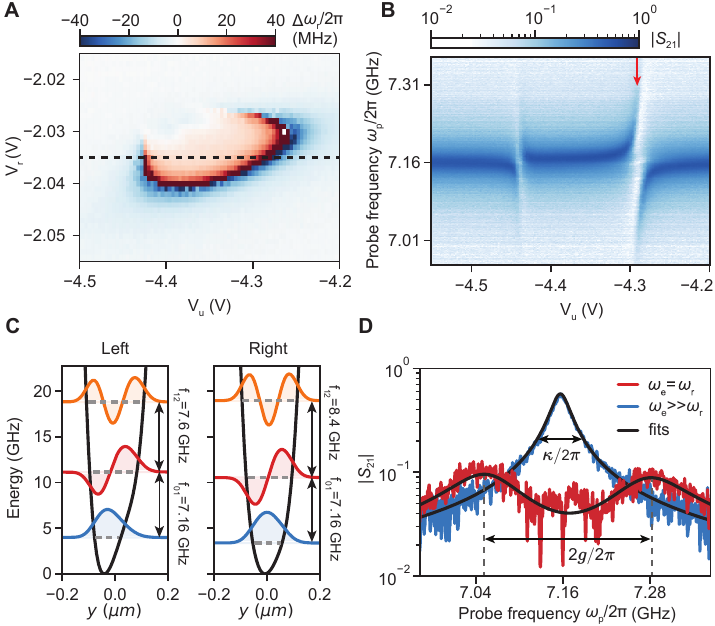}
	\caption{\textbf{Single electron strong coupling and vacuum Rabi splitting} (\textbf{A}) Resonance frequency shift $\Delta \omega_{r}/2\pi$ from a single electron versus $V_{r}$ and $V_{u}$. The experimental voltages used are found in table~\ref{tab:compensating_voltages_2d} in the Supplementary Text. (\textbf{B}) Resonator transmission amplitude $\lvert{S_{21}}\rvert$ versus $V_{u}$ at $V_{r} = -2.035$V (black dashed line in Fig.~\ref{fig:fig2}\,A), revealing two anti-crossings when $\omega_e=\omega_r$. $\lvert{S_{21}}\rvert$ is normalized to the peak transmission off-resonance and is compensated for interference with direct background transmission across the resonator. (\textbf{C}) Simulated first and second excited state wavefunctions for left and right anti-crossings at $V_{u} = -4.441$V and $V_{u} = -4.291$V using parameters from table~\ref{tab:compensating_voltages_1d}. (\textbf{D}) Resonator transmission spectrum $|S_{21}|$ (Supplementary Text, S\ref{supp:modeling_resonator_transmission_spectra}) for $\omega_{e} \gg \omega_{r}$ ($V_{u} = -4.2$V, blue trace) and $\omega_{e} \approx \omega_{r}$ ($V_{u} = -4.291$V, red trace, arrow in \textbf{(B)}). We extract $g/2\pi = 118 \pm 3$~MHz, $\Gamma_2 / 2\pi = 75 \pm 5 $~MHz and $\omega_e/2\pi = 7.169 \pm 0.004$~GHz.}
\label{fig:fig2}
\end{figure*}

To characterize the properties of the single electron cQED system, we load one electron into the dot and measure the resonator frequency shift $\Delta \omega_{r}/2\pi$ as we tune $\omega_{e}$ by adjusting the Resonator and Unload voltages (Fig.~\ref{fig:fig2}\,A). For a range of voltages, we observe large frequency shifts exceeding several resonator linewidths that appear when the electron motion approaches that of the resonator ($\omega_{e} = \omega_{r}$), which are confirmed by FEM simulations (Supplementary Text, S\ref{sec:supp_fem_adjustments}). In the upper left-hand side of the resonance contours in Fig.~\ref{fig:fig2}\,A, the measured shifts $\Delta \omega_{r}/2\pi$ are smaller than those observed for other voltage configurations. FEM indicates that this may originate from the population of higher-lying excited states of the electron motion due to relatively small trap anharmonicity for those voltages (Supplementary Text, figure~\ref{fig:supp_higher_order_transitions}). 
Fig.~\ref{fig:fig2}\,B presents the resonator transmission $|S_{21}|$ at a fixed resonator bias of $V_{r}=-2.035$~V (dashed line in Fig.~\ref{fig:fig2}\,A) while $V_{u}$ is swept, revealing two prominent anti-crossings. The transmission signal has been corrected to account for an asymmetric line shape that arises from interference with a background transmission through the resonator~\cite{Rieger2023} (Supplementary Text, S\ref{supp:modeling_resonator_transmission_spectra}). 
The two avoided crossings in Fig.~\ref{fig:fig2}\,B correspond to two distinct voltage configurations that produce a resonant interaction between the microwave resonator field and the quantized electron motional state in the dot (Fig.~\ref{fig:fig2}\,C). 



In the strong coupling regime, the coherent hybridization of electron and resonator quantum states produces two clearly distinguishable modes in the resonator transmission spectrum, also referred to as vacuum Rabi splitting~\cite{SanchezMondragon1983}. Fig.~\ref{fig:fig2}\,D shows the magnitude of the complex transmission $\left| S_{21} \right|$ as the resonator probe frequency is tuned for two values of $V_{u}$. At $V_{u}=-4.2$\,V $\omega_{e} \gg \omega_{r}$ and the spectrum is close to that of a bare resonator (blue trace). However, at $V_{u}=-4.291$~V $\omega_{e} \approx \omega_{r}$ and two distinct peaks emerge separated by the vacuum Rabi splitting $2g/2\pi$ (red trace), which unambiguously confirms that the electron is strongly coupled to the resonator. We fit the spectra in Fig.~\ref{fig:fig2}\,D to input-output theory of the coupled electron-resonator system (Supplementary Text, S\ref{supp:modeling_resonator_transmission_spectra}) and extract a coupling strength of $g/2\pi=118\pm3$~MHz and an electron decoherence rate of $\Gamma_{2}/2\pi=75\pm5$~MHz, corresponding to a single photon cooperativity $\mathcal{C} = 4g^2/\kappa\,\Gamma_{2} \simeq 32$. The experimentally extracted value of $g$ shows good agreement with the designed value, indicating the accuracy of FEM simulations in capturing the electric field distribution within the quantum dot. This contrasts with semiconductor quantum dot systems where detailed knowledge of the electric field profile is often limited due to fabrication-induced variability and nearby charge defects, making it challenging to accurately determine the electron's position~\cite{Mi2017,burkard2023Semiconductor}.


\begin{figure*} 
\centering
	\includegraphics[width=0.7\linewidth]{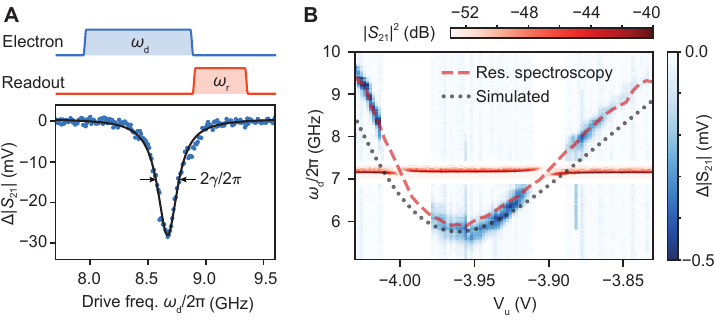}
	\caption{\textbf{Electron charge state spectroscopy} (\textbf{A}) Spectroscopy drive pulse (frequency $\omega_{d}/2\pi$, duration 50~$\mu$s) followed by resonant readout pulse (frequency $\omega_{r}$, duration 1~$\mu$s) applied to the resonator input port. A dip in readout transmission occurs at the electron frequency $\omega_{e}/2\pi~=~8.66$~GHz with a half width at half maximum $\gamma/2\pi = 102 \pm 2$ MHz~\cite{Schuster2005}. (\textbf{B}) Electron motional state spectrum (blue color map) versus $V_{u}$, showing two voltages where $\omega_{e} \approx \omega_{r}$, which agree with the avoided crossings visible in the resonator spectroscopy (red color map and red dashed curve). Simulated electron frequency based on $\Delta \omega_{r}$ (dotted grey curve) agrees with measured $\omega_{e}$ (Supplementary Text, S\ref{supp:estimating_fe_from_resonator_transmission}). Areas where $\omega_{e}/2\pi \simeq 7.162$ GHz are omitted due to fast leakage of the drive tone through the resonator.}
\label{fig:fig3}
\end{figure*}

We next directly probe the electron spectrum by performing two-tone spectroscopy~\cite{Schuster2005} using the pulse sequence in Fig.~\ref{fig:fig3}\,A. We begin by shaping the electrostatic trapping potential such that the electron motional state is far detuned from the bare resonator frequency by $\sim$1~GHz. We then apply a variable-frequency drive tone $\omega_{d}/2\pi$ to the resonator for a fixed time to excite the electron, immediately  followed by a readout tone at the resonator center frequency $\omega_{r}/2\pi$. As shown in Fig.~\ref{fig:fig3}\,A, the dispersive interaction between the electron and resonator leads to a decrease of the transmitted readout signal $\Delta \left |S_{21} \right |$ when $\omega_{d}/2\pi$ is resonant with the electron motional state transition. By fitting the resulting signal to a Lorentzian we extract the electron motional frequency $\omega_{e}/2\pi$ and two-tone linewidth $\gamma/2\pi$ which is in reasonable agreement with the values determined from resonator spectroscopy. 

To demonstrate control over the quantized electron motional states, we extract the electron frequency from two-tone spectroscopy measurements as we tune $\omega_{e}$ with $V_{u}$. In Fig.~\ref{fig:fig3}\,B we highlight the agreement between the electron frequency determined from resonator spectroscopy (dashed red curve), two-tone spectroscopy (blue data), and FEM (grey dotted curve) as we tune the trapping potential. In particular, the electron frequency we extract from resonator spectroscopy by using the dispersive approximation $\omega_{e} = \omega_{r} + g^{2}/\Delta\omega_{r}$ (Supplementary Text, S\ref{supp:estimating_fe_from_resonator_transmission}), closely follows the values determined from two-tone spectroscopy for an electron-photon coupling strength of $g/2\pi=110$~MHz over a broad range of voltages, implying that the coupling does not depend strongly on $V_{u}$. This value of $g$ is also in agreement with the independently determined value of the coupling strength from the vacuum Rabi splitting. The non-monotonic dependence of $\omega_{e}/2\pi$ on $V_{u}$ arises from the fact that the minimum of the trapping potential moves off-center while sweeping $V_{u}$ (Fig.~\ref{fig:fig2}\,C). Both an analytical model and FEM simulations suggest that our observation that the electron frequency never falls below 5.5 GHz can be explained by an in-plane electric field at the location of the electron, potentially arising from stray charges in the vicinity of the dot (Supplementary Text, S\ref{supp:analytical_model_of_electron_spectrum}).

\begin{figure*} 
\centering
	\includegraphics[width=0.7\linewidth]{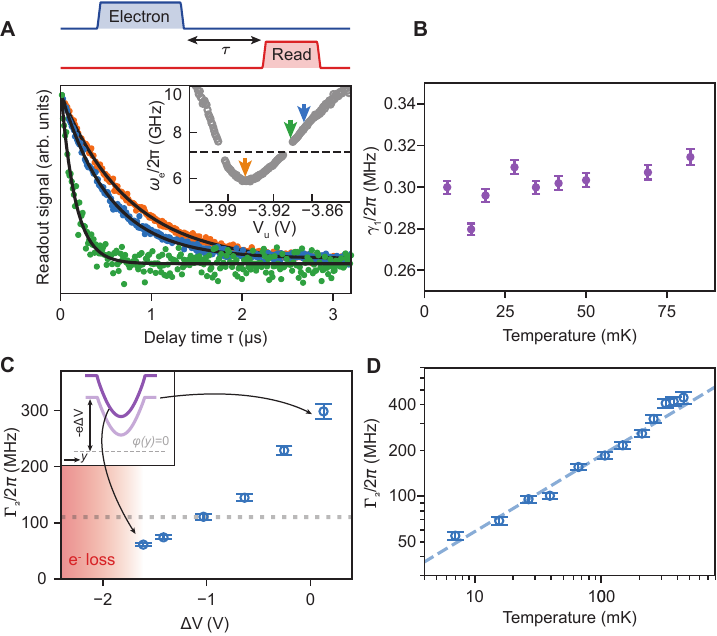}
	\caption{\textbf{Energy decay and dephasing of electron motional state} \textbf{(A)} $T_{1}$ measurements taken for three different $V_{u}$ for $\omega_{e}<\omega_{r}$ (orange), $\omega_{e}>\omega_{r}$ (blue), and $\omega_{e}\simeq \omega_{r}$ (green). The decay rate increases due to Purcell loss through the resonator when the electron-resonator detuning is small (green). Inset shows electron frequency dependence from Fig.~\ref{fig:fig2}\,B with arrows indicating electron frequency at each $T_{1}$ measurement.  (\textbf{B}) Electron decay rate $\gamma_{1}$ versus temperature, consistent with spontaneous emission. \textbf{(C}) Decoherence rate $\Gamma_{2}$ decreases when negative voltage offsets $\Delta V$ are applied to the dot electrodes. The dashed line indicates $g/2\pi=110$~MHz. Inset shows how  offset $\Delta V$ lifts the potential energy relative to the ground potential $\varphi(y)=0$. Dot electrode voltages used for each data point are provided in table~\ref{tab:lifted_dot_voltages}. While more negative offsets result in a reduction in $\Gamma_2$, there is an experimental limit beyond which the electron is no longer confined in the dot. (\textbf{D}) Decoherence rates $\Gamma_{2}$ measured at temperatures between 7~mK and 450~mK. Dashed line with slope of 0.5 in log-log scale plotted as a guide to the eye. }
\label{fig:fig4}
\end{figure*}

Two-tone spectroscopy of the electron motional state spectrum provides a stepping stone towards time domain measurements, for example to probe the electron's energy relaxation time. The electron state decoherence contains contributions from both energy relaxation and pure dephasing, i.e. $\Gamma_{2} = \gamma_{1}/2 + \gamma_{\phi}$. To determine the dominant contribution to the total linewidth we perform independent time-domain measurements to extract the electron energy relaxation time, $T_{1} = 1/\gamma_{1}$ as shown in Fig.~\ref{fig:fig4}\,A. We apply a $20\,\mu$s pulse at frequency $\omega_{e}/2\pi$ followed by a variable length delay $\tau$ and subsequent readout pulse at $\omega_{r}/2\pi$. As $\tau$ increases, the transmitted readout signal decays exponentially with a characteristic time constant $T_{1}$, which we extract from fitting. The data in Fig.~\ref{fig:fig4}\,A are taken for three different electron frequencies corresponding to differing values of the electron-resonator frequency detuning (blue, green, and orange). At large electron-resonator detuning we find a maximum $T_1 \simeq 0.76~\mu$s, while for smaller values of detuning energy relaxation is enhanced, likely from Purcell decay into the resonator (Supplementary Text, S\ref{supp:purcell_decay}). Additionally, as shown in Fig.~\ref{fig:fig4}\,B, we find that $\gamma_{1}$ is temperature independent, consistent with spontaneous emission. Importantly these measurements show that $\gamma_1$ is two orders of magnitude smaller than $\Gamma_2$, indicating that the electron decoherence is primarily limited by pure dephasing, not energy relaxation. 

Previous attempts at reaching the strong coupling regime with electrons on helium suggested that low-frequency classical helium surface fluctuations could be a dominant decoherence mechanism~\cite{Koolstra2019}. These surface fluctuations, arising from mechanical vibrations in the system, can in principle modulate the electron’s out-of-plane position, leading to variations in the in-plane trapping potential. This can alter the electron frequency and potentially cause dephasing. However, we observed no significant change in the two-tone linewidth $\gamma$ when the level of mechanical vibration of our cryostat was increased by an order of magnitude by disabling the active vibration cancellation system (Supplementary Text, S\ref{supp:classical_helium_vibrations}).

To further investigate what limits $\Gamma_{2}$, we study the effect of modifying the trapping potential on the electron decoherence rate by applying a negative offset $\Delta V$ to the trapping electrodes relative to the nearby ground electrode (Fig.~\ref{fig:fig4}\,C). Notably, a transition from the weak to strong coupling regime is observed as the offset is made progressively more negative, with the lowest measured decoherence rate reaching $\Gamma_{2}/2\pi=61$ MHz. This transition is unambiguously confirmed by the appearance of a vacuum Rabi splitting in the measured resonator spectroscopy for voltage configurations with offset $\Delta V < -1$~V (Supplementary Text, S\ref{supp:temp_dependent_rabi_splitting}). The observed decrease in $\Gamma_2$ may be attributed to a reduced electron dephasing as the electric field pressing the electron against the helium surface becomes weaker. This behavior is expected for electrons interacting with the bosonic field of quantized surface capillary waves (ripplons)~\cite{Dykman2023}. Alternately, the voltage offset applied to the electrodes may change the spatial distribution of nearby stray electrons that are potentially introduced during the initial electron emission process and that couple to the trapped electron. Applying more negative $\Delta V$ could repel these fluctuating stray charges (Fig.~\ref{fig:fig4}\,C), thereby reducing dephasing from charge noise, a ubiquitous decoherence mechanism for solid state qubits~\cite{RojasArias2023, Li2025}.

We further demonstrate that the decoherence rate has a strong temperature dependence, as shown in Fig.~\ref{fig:fig4}\,D, changing almost tenfold from 7~mK to 450~mK (Supplementary Text, S\ref{supp:temp_dependent_rabi_splitting}). The value of $\Gamma_2$ closely follows a power-law dependence $T^\beta$ with $\beta \approx 0.5$, and shows no sign of saturation at the lowest temperatures. This behavior is consistent with the coupling of the electron state to a thermally activated bath having low-frequency excitations~\cite{Krantz2019,Dykman2023}, which could originate from either the ripplonic field of the helium surface or the fluctuation of stray charges in the vicinity of the dot.

In conclusion, our result of strong coupling between the electron motional state on helium and the microwave resonator photons opens access to the investigation of a range of new light-matter phenomena with a single fundamental particle. Additionally, our demonstration of strong coupling is a vital ingredient for coupling an electron spin on helium to a microwave photon using well-established cQED methods similar to those used in semiconductor quantum dots~\cite{Samkharadze2018, Mi2018}. Notably, previous demonstrations with similar charge decoherence and coupling have achieved spin readout via charge state hybridization~\cite{Samkharadze2018}, placing microsecond spin qubit readout on helium within reach (Supplementary Text, S\ref{supp:prospects_for_spin_coupling}).



\newpage

\bibliography{thebibliography} 

\newpage

\renewcommand{\thefigure}{S\arabic{figure}}
\renewcommand{\thetable}{S\arabic{table}}
\renewcommand{\theequation}{S\arabic{equation}}
\renewcommand{\thepage}{S\arabic{page}}
\setcounter{figure}{0}
\setcounter{table}{0}
\setcounter{equation}{0}
\setcounter{page}{1} 



\clearpage
\newpage
\clearpage
\onecolumngrid

\section*{\large Supplementary Materials for\\ Strong coupling of a microwave photon to an electron on helium}




\twocolumngrid

\subsection*{Materials and Methods}

\subsubsection*{Device fabrication}
The single electron devices are fabricated on undoped high resitivity Si wafers ($>$ 10 k$\Omega \cdot$cm) with (100) orientation. Immediately after a standard solvent cleaning of the wafer, a 12 nm thick TiN film is grown by Plasma-Enhanced ALD (Ultratech/Cambridge Fiji G2) at a temperature of 270 $^{\circ}$C and 150 cycles. This superconducting film is used to fabricate the resonator, electrodes, and ground plane in our devices.

\begin{figure*}
    \centering
    \includegraphics[width=1.0\textwidth]{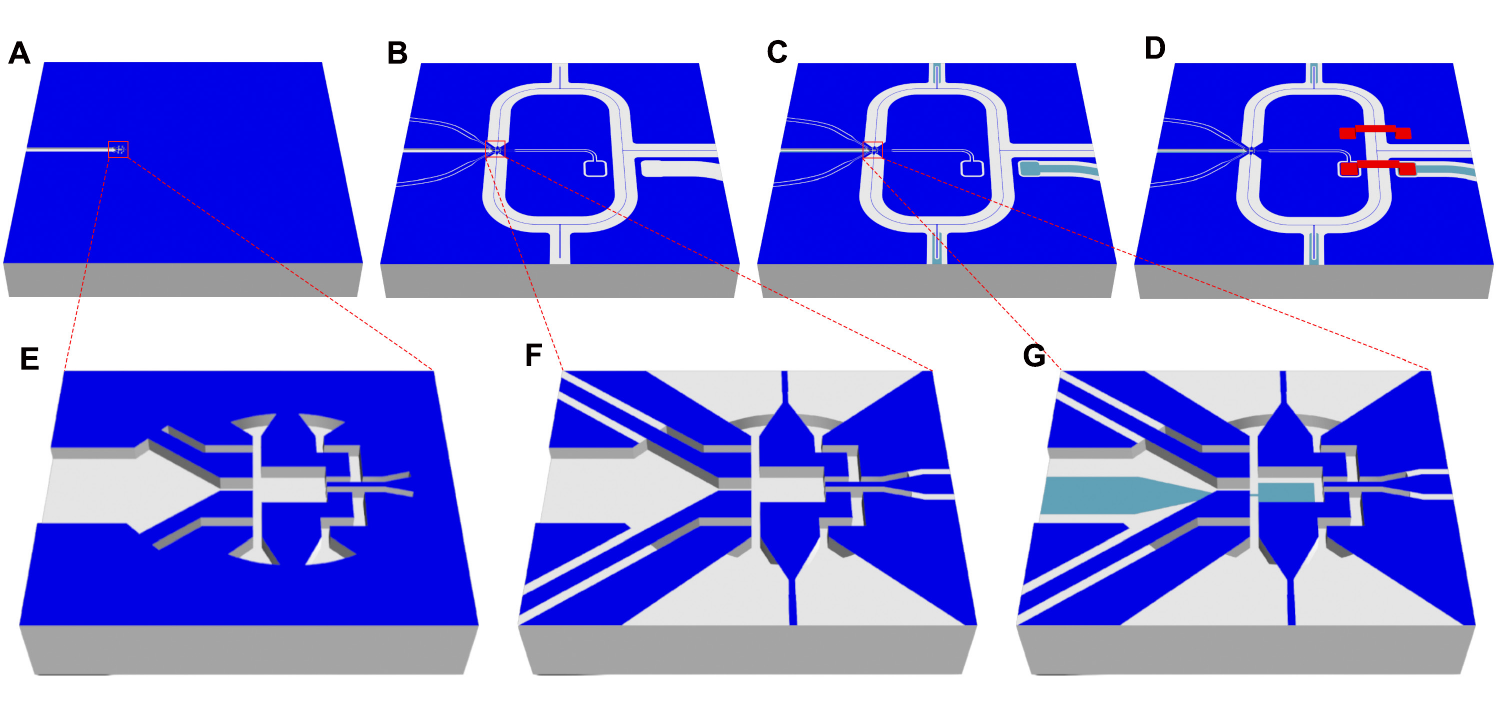}
    \caption{\textbf{Process flow of the device fabrication} \textbf{(A)} Si trenches with TiN ground plane in the electron dot region. \textbf{(B)} Patterning CPW resonator, guard electrodes, and unload electrode on TiN ground plane. \textbf{(C)} Adding Nb electrodes at the bottom of the Si trench and in TiN etched region. \textbf{(D)} Adding Al airbridges (red) to make the connection between TiN unload electrode (blue) and Nb DC bias line (light blue), as well as connection between two separate TiN ground plane. \textbf{(E)}, \textbf{(F)}, \textbf{(G)} represent the zoom-in images of the \textbf{(A)}, \textbf{(B)}, \textbf{(C)}.}
    \label{fig:fab_process}
\end{figure*}

After the TiN film growth, the geometry of the electron reservoir, the microchannel, and the electron dot is defined using a combination of e-beam lithography and dry etching. The e-beam lithography is performed with a Raith EBPG5000+ to form a resist etch mask on the TiN film surface. Subsequently, we etch Si/TiN at once using fluorine chemistry in an ICP plasma etcher. The dry etching recipe uses a gas mixture of SF$_6$ 20~sccm : C$_4$F$_8$ 30~sccm, ICP power 600~W, and RF bias power 60~W at 10~$^{\circ}$C and 10~mTorr pressure. To remove the remaining resist after etching, the sample is soaked in AZ NMP rinse at 80~$^\circ$C overnight, followed by 30~min of sonication with a fresh AZ NMP rinse at 80~$^\circ$C and an IPA rinse at the end. This process forms $\sim$0.78 $\mu$m deep TiN covered Si trenches (see figure~\ref{fig:fab_process}\,A), where liquid helium is filled up by capillary action in the experiment.

The next patterning step is to build the resonator, the Guard, and the Unload electrodes by dry etching TiN with a resist etch mask, as shown in figure~\ref{fig:fab_process}\,B. The etch mask is first patterned on the sample using e-beam lithography (Raith EBPG5000+) and then the area of TiN not protected by the mask is etched away using the same recipe and the same tool as above. The total etching time of TiN is extended by $\sim$50~$\%$ to ensure there is no TiN residue, leading to additional Si etching by $\sim$90~nm. After etching TiN, the resist mask is first partially removed by an oxygen plasma using a downstream plasma asher YES-CV200 RFS, followed by putting the sample in an AZ NMP rinse at 80~$^\circ$C overnight. The sample is sonicated with fresh NMP at 80~$^\circ$C for 1.5~hours and rinsed with IPA at the end. 

To add additional electrodes, we perform another patterning process using a lift-off method. First, an e-beam resist pattern is formed with another round of e-beam lithography, followed by a 80 nm thick Nb layer deposition using a Plassys MEB550S Electron Beam Evaporator. For resist removal after Nb deposition, the sample is soaked in AZ NMP rinse at 80~$^\circ$C overnight, and then sonicated with a fresh AZ NMP rinse at 80~$^\circ$C for 20~min. The remaining NMP is cleaned away by an IPA rinse. As a result, the Bottom electrode in the dot region is made of Nb situated at the bottom of the $\sim$0.78 $\mu$m deep TiN-covered Si trench (light blue electrode in figure~\ref{fig:fab_process}\,G). The resonator feedline and DC bias lines on the TiN ground plane are formed of Nb placed in the $\sim$100~nm deep TiN etched region (light blue electrodes in figure~\ref{fig:fab_process}\,C). The Guard and Unload electrodes (TiN) are electrically connected to their own Nb DC bias lines through the Al airbridges, of which the process is described below.

The fabrication process for the Al airbridges is adapted from Ref.~\cite{Chen2014}. Using optical lithography, we first form the scaffold to support the airbridges with 3~$\mu$m thick positive photoresist. The resist is then re-flowed at 140~$^\circ$C for better mechanical stability. Next, 300~nm thick aluminum is deposited at high-vacuum using a Plassys MEB550S Electron Beam Evaporator to build the metal layer of the airbridge. Before the aluminum deposition, in-situ ion milling is performed to remove the native oxide on the TiN and Nb surface for better ohmic contact with the aluminum. Through another round of optical lithography, the bridge pattern is defined with the same thick photoresist and the aluminum not used for the airbridge is removed by wet etching with Transene Aluminum Etchant Type A. Before aluminum wet etching, the resist is baked at 115~$^\circ$C to improve adhesion to the aluminum. After wet etching, the resist is removed by a combination of dry and wet resist strip processes. During the dry strip process, most of the cross-linked resist formed during the ion milling step prior to the aluminum deposition is removed by oxygen plasma. The sample is then soaked in AZ NMP rinse to remove all remaining resist, followed by rinsing with IPA. Figure~\ref{fig:fab_process}\,D shows the Al airbridge (red) used to connect the TiN Unload electrode (blue) to the Nb DC bias line (light blue). The TiN Guard electrodes are also connected to their Nb DC bias lines in the same way, which is not shown in figure~\ref{fig:fab_process}. In addition to electrodes connection, the Al airbirdges are also put over the most of the signal lines to connect the two TiN ground planes on either side of the lines for slotline modes suppression.

\subsubsection*{Sample cell and helium filling}

\begin{figure*}
    \centering
    \includegraphics[width=0.5\textwidth]{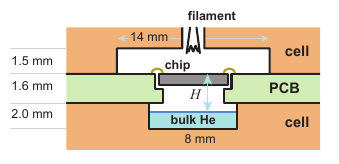}
    \caption{\textbf{Device enclosure} Cross-sectional schematic view of the chip integrated within the hermetically sealed sample enclosure indicating the bulk helium reservoir region and tungsten filament used for electron emission.}
    \label{fig:cell_crossection}
\end{figure*}

The fabricated chip of size 2~mm$~\times~$7~mm is wirebonded to the host PCB which is hermetically sealed within the sample cell, as schematically illustrated in figure~\ref{fig:cell_crossection}. An additional volume  beneath the chip serves as a bulk liquid helium reservoir. The top section of the cell is connected to a large 150~cc reservoir used to store helium gas prior to cooldown. A small tungsten filament positioned several millimeters above the chip serves as the source of electrons.

The $^4$He is isotopically enriched and sourced from Lancaster Helium Ltd. We pressurize the sample cell at room temperature to 0.9~bar, corresponding to a bulk liquid helium volume of approximately 0.18~cm$^3$. A liquid nitrogen cold trap between the gas cylinder and the experimental cell is used to filter impurities such as water and nitrogen. Immediately before cooling down, we seal the experimental cell using the method outlined in \cite{Castoria2024}.

Below the superfluid transition temperature, liquid helium fills the on-chip microchannels via capillary action, with a radius of curvature set by the distance from the chip to the bulk helium reservoir $H$ (see figure~\ref{fig:cell_crossection}). From the sample cell geometry we estimate $H \approx 3$~mm, which results in a depression of the helium surface at the center of the dot of approximately \cite{KoolstraThesis, Marty1986} $\rho_\mathrm{He} G H w_\mathrm{ch}^2 / 8 \sigma_\mathrm{He} \approx 2$~nm. Here, $\rho_\mathrm{He}$, $\sigma_\mathrm{He}$ are the density and surface tension of superfluid helium, respectively, $w_\mathrm{ch} = 1.4$~$\mu$m is the width of the electron dot, and $G=9.81$~m/s$^2$ is the earth's gravitational acceleration.

\subsubsection*{Cryogenic setup}
\begin{figure*}
    \centering
    \includegraphics[width=0.95\textwidth]{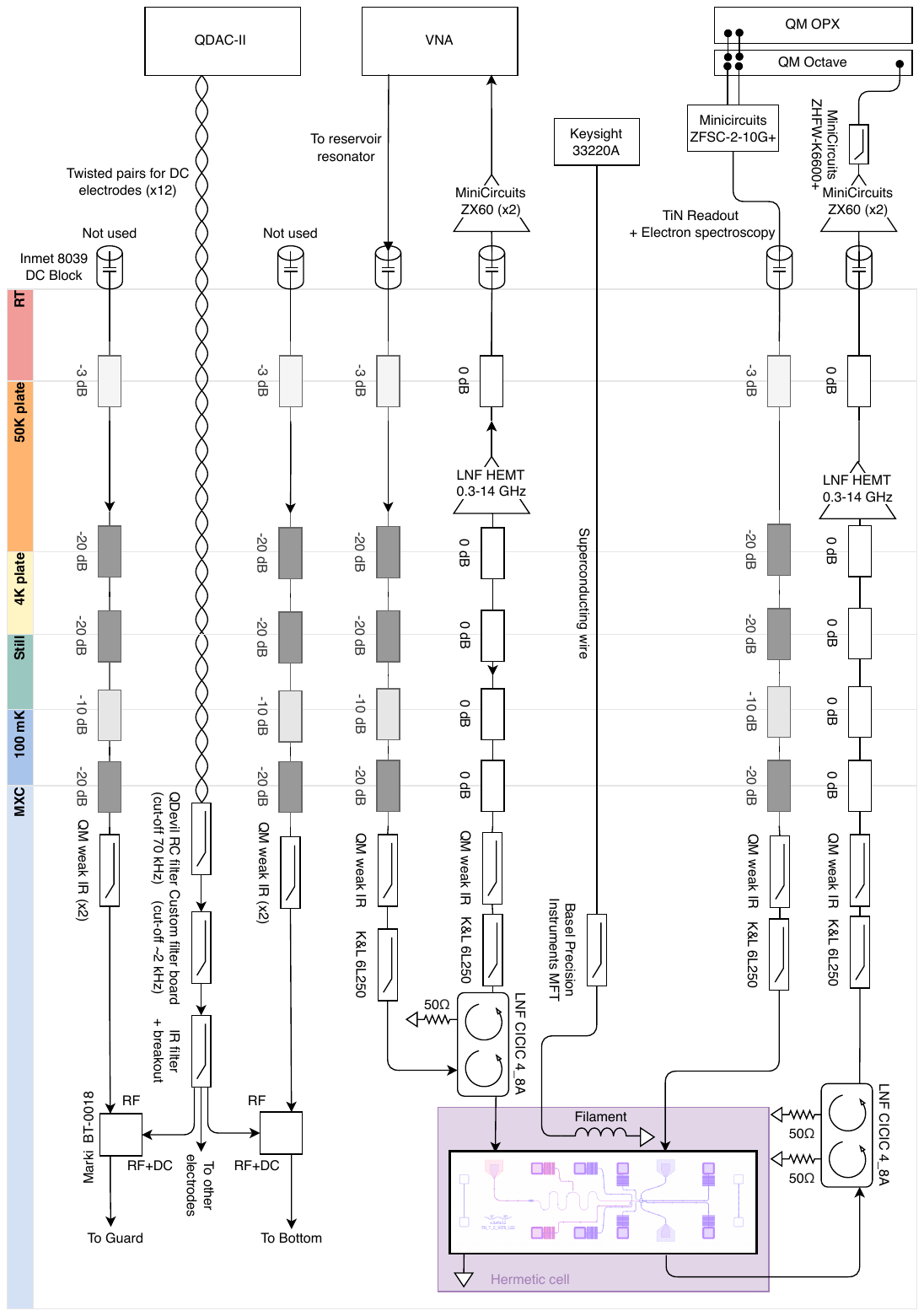}
    \caption{\textbf{Dilution refrigerator wiring diagram and experimental instrumentation.} Refrigerator stages are color coded from room temperature (red, RT) to the mixing chamber plate (blue, MXC). The device is located inside a hermetic sample enclosure. For complete details see Materials and Methods.}
    \label{fig:fig_s1}
\end{figure*}

All measurements were performed in a Bluefors LD400 dilution refrigerator with a base temperature of 7~mK. The experimental configuration for electron spectroscopy measurements of Fig.~\ref{fig:fig3} and \ref{fig:fig4} is shown in figure~\ref{fig:fig_s1}. 



Electrode DC voltages are supplied by a QDevil QDAC II which connects to the Bluefors twisted pair loom. We filter DC lines at the mixing chamber stage, with a stack consisting of a QDevil RC filter, a custom built filter board containing several $\pi$-filters and lastly, a filter box filled with infrared absorber to filter frequencies above 1~GHz. We measure a cut-off frequency for this filter stack of approximately 2~kHz, which still allows for fast voltage sweeps if needed. The outputs of the filter stack go directly to the hermetic cell, except for one of the Guard electrodes and the Bottom electrode, which instead pass through a bias tee. The RF port of both bias tees is not used in this experiment.

\subsubsection*{Electron reservoir}
Electrons are emitted from the tungsten filament located inside the hermetic sample cell. One terminal is connected to an Keysight 33220A arbitrary waveform generator (configured in single pulse mode) while the other end is grounded. We typically observe electron emission by pulsing the filament for a duration of 100~ms and amplitude of -2.4~V, while the cell is thermalized at a temperature of 0.65~K. Electrons are then captured by applying a positive voltage (typ. 1.5~V) to the center pin of a niobium $\lambda/2$-resonator ($\omega_{r}/2\pi = 6.527$~GHz), located on the bottom of a 0.7~$\mu$m-deep micro-channel \cite{Yang2016}. This results in a dispersive resonance frequency shift of several MHz signaling the presence of electrons on top of the reservoir resonator. Because the micro-channel connects to the dot shown in Fig.~\ref{fig:fig1}\,B, electrons can now be transferred from the reservoir resonator to the dot with the appropriate voltages.

To transfer electrons into the dot, we bias the Bottom electrode at 0.4~V and the Guard electrodes close to 0.0~V, and decrease the voltage on the reservoir resonator until a negative frequency shift is observed on the TiN resonator. During experiments, the reservoir is kept at 2.0~V such that any stray electrons close to the dot are drained to the reservoir resonator.

Electron readout and spectroscopy pulses are generated by a Quantum Machines OPX and Octave unit. Both readout and spectroscopy pulses are combined at the TiN readout resonator input line, are attenuated by approx. -73~dB and then filtered (Quantum Microwave IR filter and K\&L~6L250 low-pass filter). The transmission is filtered and amplified by an LNF high-electron mobility transistor (HEMT) amplifier, and further amplified at room temperature before the signal is processed by the Octave.

We probe the electron reservoir resonator with a vector network analyzer (VNA, Keysight E5071C) and measure the resonator in reflection ($S_{11}$) using a double junction circulator (LNF CICIC 4\_8A) mounted at the mixing chamber. Except for the final high-pass filter, the return chain for the reservoir resonator is identical to the return chain of the TiN resonator.

\subsubsection*{Finite Element Modeling (FEM) of trapping potential}
The behavior of both the many-electron system and a single electron within the dot region is entirely governed by the local electrostatic potential at the helium surface. This potential is controlled via the voltages $V_i$ applied to the electrodes defining the dot area and is expressed as $\varphi(x, y) = \sum_{i} \alpha_i(x, y)\, V_i$, where $\alpha_i(x, y)$ represents the relative contribution of the $i$-th electrode to the overall potential. We employ Finite Element Methods (FEM) to compute $\alpha_i(x, y)$ by setting each $i$-th electrode to 1~V while grounding all other electrodes at 0~V, repeating this procedure for every electrode in the system~\cite{zhk,Castoria2025}. The resulting potential is then used to determine the many-electron ground-state configuration and their vibrational frequencies. The single-electron excitation spectrum is obtained by numerically solving the two-dimensional Schrödinger equation for an electron in a trapping potential $\varphi(x, y)$, from which we extract the electron frequencies $\omega_{e}$ and anharmonicities $\alpha_e$. We also compute the microwave field distribution produced by the resonator pins and use it to estimate $\ell = \big (\partial\alpha^{-}/\partial y)^{-1}$. Here $\alpha^{-}$ is obtained by setting upper and lower arms of the resonator to $+1/2~V$ and $-1/2$~V respectively and all other electrodes are kept at zero potential. This estimate is then used to determine the electron–photon coupling strength.

\clearpage
\newpage
\clearpage
\onecolumngrid


\section*{\large Supplementary Text}

\twocolumngrid

\section{Electron-photon coupling energy}
\label{supp:g}

The coupling energy between the single electron motional state and the resonator photon is given by $\hbar g = |\vec{d}|\,V_{\mathrm{zpf}}/\ell$, where $\vec{d} = -e l_y \vec{\mathbf{y}}$ is the dipole moment of the electron, $l_y = \sqrt{\hbar/2m_e\omega_e} \approx 36$~nm is the zero-point motion amplitude of electron in the $\vec{\mathbf{y}}$-direction (on resonance $\omega_e = \omega_r$), $V_{\mathrm{zpf}}$ is the zero point voltage fluctuation of the resonator and $\ell$ parametrizes the effective size of the region over which the microwave field is concentrated within the dot. We design a compact quantum dot, resulting in $\ell \approx 2.2$~$\mu$m at its center, as determined from FEM calculations. This value is slightly larger than the physical size of the dot due to electric field screening effects from surrounding electrodes (primarily the Bottom electrode), which partially shield the field generated by the resonator electrodes. 

Leveraging the high kinetic-inductance of TiN, we design a high-impedance resonator composed of two arms formed by coplanar waveguide structures. Each arm has a center nanowire with a width of 175~nm and a gap of 5~$\mu$m to the adjacent ground plane. Each arm of the resonator has a length of 135~$\mu$m. At the open end, the resonator arms are terminated at the quantum dot, which is modeled as a capacitive coupler between two arms with capacitance $C_\mathrm{dot} = 0.028\ \text{fF}$, as determined from FEM. At the opposite end, the two arms are galvanically connected to each other and to an additional nanowire (``tail"), which allows a DC bias to be applied to the nanowire. We model the nanowire resonator as two symmetrically coupled lumped LC-circuits~\cite{Koolstra2025,Castoria2025}, where each half has capacitance $C_r = 5.8$~fF and inductance $L_r = 85$~nH. This value is within 10\% of the value calculated from a conformal mapping calculation. Our resonator supports two modes distinguished by their spatial voltage distributions: the \textit{differential} mode, characterized by out-of-phase voltage fluctuations at the open ends of the arms, and the \textit{common} mode, which has in-phase voltage fluctuations~\cite{Castoria2025}. Due to the symmetry of the quantum dot geometry, only the differential mode produces an electric field capable of driving electron motion. The estimated resonator impedance for the differential mode is $Z = \sqrt{L_r/C_r} \approx 3.8\ \text{k}\Omega$, corresponding to a zero-point voltage fluctuation amplitude of $V_{\mathrm{zpf}} = \sqrt{2\hbar\omega_r/C_r} \approx 40\ \mu\text{V}$. This yields a designed single-electron coupling strength of $g/2\pi \approx 110\ \text{MHz}$. 

In the experiments the voltages are tuned to align the electron motion with the electric field of the resonator’s differential mode along the $\vec{\mathbf{y}}$-direction, thereby maximizing the coupling between the electron and resonator photons.

In our numerical calculations described below, we set the resonator tail inductance to $L_\mathrm{tail}~=~109$~nH to match the common mode frequency with the experimentally observed value of 3.8~GHz.

\section{Electron cluster properties up to $N = 4$}
\label{supp:electron_cluster_properties}
In this section, we present the details for modeling resonance frequency shifts $\Delta \omega_{r}$ shown in Fig.~\ref{fig:fig1}\,E. The results provide an explanation for the smaller $\Delta \omega_{r}$ observed for $N = 1, \ldots, 4$ electrons shown in Fig.~\ref{fig:fig1}\,E compared to the single electron frequency shifts presented in Fig.~\ref{fig:fig2}. This behavior arises because the motional frequencies of the electron clusters are far higher than the resonator frequency ($>$20~GHz) for the voltages in Fig.\ref{fig:fig1}\,E. 

To model the resonance frequency shift due to $N$ trapped electrons, we begin by extracting the electrostatic potential from FEM simulations and numerically solve for the electron configuration that minimizes the total potential energy, taking into account the repulsive Coulomb interaction between electrons (see Fig.~\ref{fig:fig1}\,D)~\cite{Yang2016}. This is followed by formulating the coupled equations of motion in matrix form, which includes the resonator and two-dimensional electron motion, including charge-photon coupling (details in Ref.~\cite{Koolstra2025}). We numerically solve for the coupled eigenmodes and eigenfrequencies using the \texttt{quantum\_electron} python package~\cite{Koolstra_Quantum_Electron_2024}. Finally, the resonator frequency shift is calculated by subtracting the bare differential mode frequency, 
\begin{align}
    \omega_{r} = \frac{1}{\sqrt{L_r C_r}},
\end{align}
from the frequency of the eigenmode with the strongest cavity participation. The resulting shifts are shown as black horizontal lines in Fig.~\ref{fig:fig1}\,E.

\begin{figure*}
    \centering
    \includegraphics[width=0.8\textwidth]{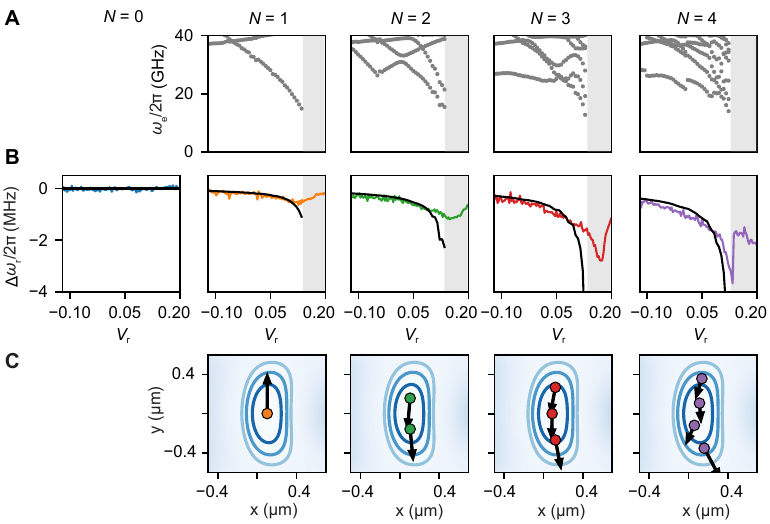}
    \caption{\textbf{Resonator voltage dependence for electron clusters of various sizes} \textbf{(A)} Top row: in-plane motion frequencies for electron clusters consisting of $N=1$ to $N=4$ electrons. Abrupt changes in the electron frequency $\omega_{e}$ can be observed due to rearrangement of the electrons within a cluster. \textbf{(B)} Bottom row: measured (colored data) and modeled resonance frequency shift (black lines) show good agreement. The gray shaded voltage regions indicate that the simulated potential can no longer hold the electron cluster, potentially explaining the kinks in the measured resonator frequency shift $\Delta \omega_{r}$. \textbf{(C)} Electron configurations at $V_{r} = 0.08$~V for $N=1$ through $N = 4$ with strongest coupled eigenvector (black arrows) pointing predominantly in the $y$-direction.}
    \label{fig:unloading_further_investigated}
\end{figure*}

\begin{figure*}
    \centering
    \includegraphics[width=0.45\textwidth]{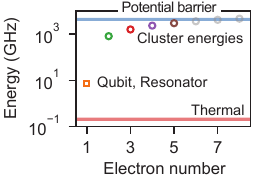}
    \caption{\textbf{Relative energy scales of the dot-confined electron system} The depth of the electrostatic trap of the quantum dot ($\sim10^3$~GHz) exceeds both the electron motional state frequency ($\simeq$4-8~GHz) of N $\leq$ 4 electrons, the resonator energy, and the thermal energy ($\simeq$0.2~GHz) by several orders of magnitude. Therefore, quantum tunneling does not play a role and electron confinement times are practically infinite.}
    \label{fig:supp:energy_scales}
\end{figure*}

Figure~\ref{fig:unloading_further_investigated}\,B shows good agreement between the modeled and measured resonance frequency shift, especially for $V_{r} < 0.12$~V. For more positive $V_{r}$ (gray shaded areas in figure~\ref{fig:unloading_further_investigated}), the total energy of the electron clusters exceeds the depth of the trapping potential (figure~\ref{fig:supp:energy_scales}), potentially explaining the kink in the measured value of $\Delta \omega_{r}$ for $N=1, \ldots, 4$. Furthermore, figure~\ref{fig:unloading_further_investigated}\,A shows that the eigenfrequency of the electron modes remains above 20~GHz for all values of $V_{r}$, which results in small measured frequency shifts $\Delta \omega_r/2\pi \leq 2$~MHz presented in Fig.~\ref{fig:fig1}\,E. These results show that the classical equations of motion accurately model the electron dynamics in the experiment.  

\section{Electron frequency calculations and FEM parameters} \label{sec:supp_fem_adjustments}
\begin{figure*}
    \centering
    \includegraphics[]{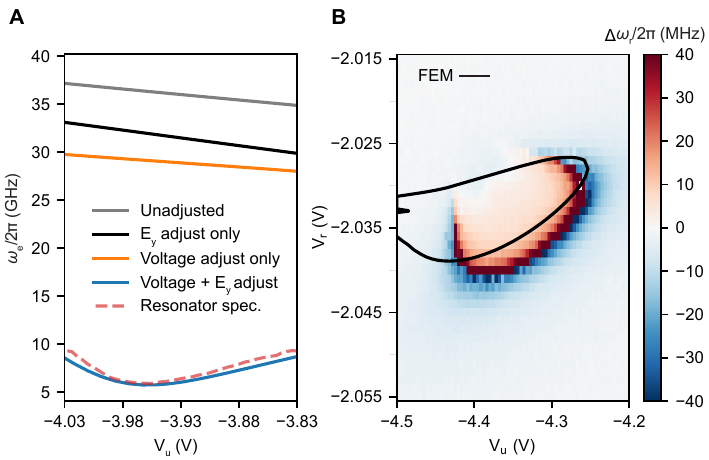}
    \caption{\textbf{Electron motional transitions extracted from FEM (A)} Simulated electron frequencies for experimental voltages and adjusted FEM parameters. Adjustment includes both changes to the dot electrode voltages, and inclusion of a compensating field $E_{y}$, likely due to nearby stray charges. Best agreement between simulated values and experimental resonator spectroscopy (``Resonator spec.") found when both voltage adjustments and a compensating field are included (``Voltage + $E_{y}$ adjust"). Adjusted parameters found in table~\ref{tab:compensating_voltages_1d}. Electron frequency values are simulated by solving the 2D Schr\"odinger equation for the electrostatic potential. \textbf{(B)} Experimental resonator spectroscopy data from Fig.~\ref{fig:fig2}\,B overlaid with FEM simulation of electron resonance condition. Black line indicates where FEM simulation predicts electron frequency to be equal to resonator frequency at 7.162~GHz. Adjusted parameters used to produce black line are provided in table~\ref{tab:compensating_voltages_2d}.} 
    \label{fig:supp_voltage_comp}
\end{figure*}

Finite element modeling (FEM) of the trapping potential is carried out using the technique outlined in the Materials and Methods section. Electron frequencies are extracted by solving the single electron 2D Schr\"odinger equation using the electrostatic trapping potential from a given dot voltage configuration. In several experiments, achieving the electron resonance condition was only possible when voltage asymmetries were applied to the Guard electrodes. This suggests the presence of a stray in-plane electric field at the location of the electron in the dot. In simulations, we account for this asymmetry by adding two terms in the FEM model to represent static compensating electric fields $E_{y}$ and $E_{x}$. In addition to introducing compensating fields, adjustments are made to the other electrode voltages from their experimental values until the electron frequencies calculated from FEM simulations align well with those observed in experiments (figure~\ref{fig:supp_voltage_comp}). 
Notably, a Bottom electrode offset of $\sim +0.1$~V is needed in all adjusted voltage configurations to find good agreement between experimental and predicted FEM values. The $+0.1$~V adjustment is consistent with the difference in the metal work function of the niobium Bottom electrode and the top titanium nitride electrodes~\cite{Zhuang2022, Michaelson1977}. Other deviations in the FEM parameters from the voltages used in the experiments likely arise due to fabrication abnormalities that slightly alter the dot electrode geometry from the simulated geometry, or due to screening from nearby charges. The FEM parameters used to generate the electron frequencies in figure~\ref{fig:supp_voltage_comp}A and figure~\ref{fig:supp_voltage_comp}B are listed in table~\ref{tab:compensating_voltages_1d} and table~\ref{tab:compensating_voltages_2d}, respectively.  

\begin{figure*}
    \centering
    \includegraphics[]{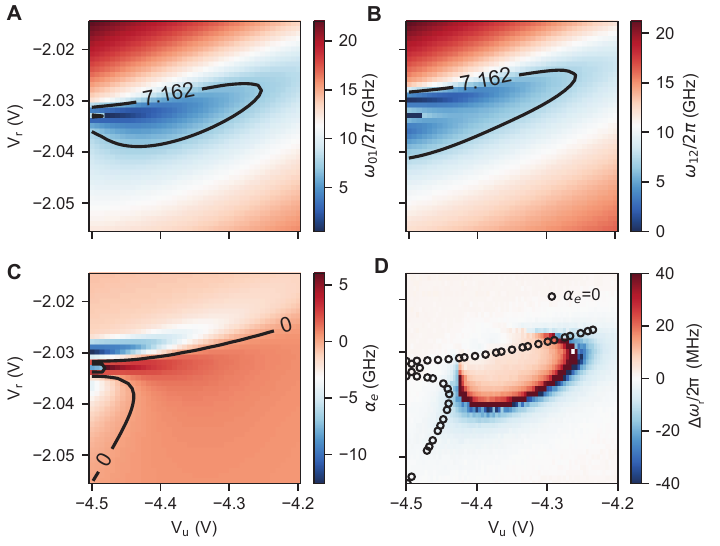}
    \caption{\textbf{FEM modeling of higher order motional state transitions} \textbf{(A)} FEM simulation of ground to first excited motional state transition frequency $\omega_{01}/2\pi$. Contour indicates where $\omega_{01}/2\pi$= 7.162~GHz. \textbf{(B)} FEM simulation of first to second excited motional state transition frequency $\omega_{12}/2\pi$. Contour indicates where $\omega_{12}/2\pi$= 7.162 GHz. \textbf{(C)} Anharmonicity of the electron excited states $\alpha_{e} = \omega_{12}/2\pi- \omega_{01}/2\pi$. Contour indicates where $\alpha_{e} = 0$ and the two transitions are degenerate. \textbf{(D)} Experimental resonator spectroscopy data shown in Fig.~\ref{fig:fig2}\,B overlaid with contours from (A) and (C) where $\omega_{01}/2\pi$ = 7.162~GHz and $\alpha_{e} = 0$, respectively. Presence of additional features in the spectra close to $V_{r}$ = -2.03~V agree well with where the two contours overlap, suggesting these features arise from coupling to higher order excited state transitions. All simulations performed using the parameters in table~\ref{tab:compensating_voltages_2d} to solve the 2D Sch\"odinger equation and calculate the quantized electron energies.}
    \label{fig:supp_higher_order_transitions}
\end{figure*}

The disappearance of strong avoided crossing features (i.e. large frequency shifts $\Delta \omega_{r}$) in the upper left region of the $\omega_r/2\pi$ map shown in Fig.~\ref{fig:fig2}A and figure~\ref{fig:supp_voltage_comp}B indicate the deviation of the electron susceptibility from a two-level model and may be attributed to excitations into higher-lying excited states. This behavior is qualitatively supported by our FEM simulation results. In figure~\ref{fig:supp_higher_order_transitions} we show the transition frequency between the ground and first excited states $\omega_{01}$~(A), along with transition frequency between the first and second excited states $\omega_{12}$~(B) and the electron anharmonicity $\alpha_{e}=(\omega_{12}-\omega_{01})/2\pi$~(C) calculated using the voltages in table~\ref{tab:compensating_voltages_2d}. We find that for the voltage configurations $V_u < -4.45$~V and $-2.04$~V~$< V_r < $~$-2.03$~V both $\omega_{01} = \omega_r$ and $\omega_{12} =\omega_r$ contour lines exhibit complex shapes. In combination with observation that $\alpha_e = 0$ in the same region, this might result in a nontrivial and potentially complex electron response. 


For the voltage configurations at which we perform the measurements in Fig.~\ref{fig:fig2}B, Fig.~\ref{fig:fig2}D, Fig.~\ref{fig:fig3}, and Fig.~\ref{fig:fig4} the predicted anharmonicities are large, and allow us to treat our electron as a two level system.

\section{Modeling resonator transmission spectra}
\label{supp:modeling_resonator_transmission_spectra}
To model the experimental transmission spectrum (including effects of coupling to the electron), we start with the following Hamiltonian in the frame rotating with a probe tone at frequency $\omega_p/2\pi$~\cite{Bonsen2023}: 
\begin{align}
\mathcal{H}/\hbar = &(\omega_r - \omega_p) a^\dagger a + (\omega_e - \omega_p) \frac{\sigma_z}{2} \notag \\
&+ g (a^\dagger \sigma^- + a \sigma^+) + i \sqrt{\kappa_\mathrm{1}} (a_\mathrm{in, 1} a^\dagger - a_\mathrm{in, 1}^* a).
\end{align}
Here, $\omega_r/2\pi$ is the bare resonator frequency, $\omega_e$ is the bare electron frequency, $a^\dagger, a$ are the raising and lowering operators of the resonator, $\sigma^\pm = \sigma_x \pm i \sigma_y$ are the Pauli matrices raising and lowering a electron excitation respectively. Further, $a_\mathrm{in, 1}$ is the photon flux of the input field (applied at port 1 of the resonator) and $\kappa_\mathrm{1, 2}/2\pi$ are the input and output coupling rates for the resonator. Note that this Hamiltonian approximates the electron as a two level system. 

To calculate transmission spectra from this Hamiltonian we use the Heisenberg equations of motion for $a$ and $\sigma^-$, which read
\begin{align}
    &\dot{a} = i (\omega_r - \omega_p) [a^\dagger a , a] + i g \sigma^- \left[ a^\dagger , a \right] - a_\mathrm{in, 1} \sqrt{\kappa_1} \left[ a^\dagger , a \right] \\
    & \dot{\sigma}^- = \frac{i}{2} (\omega_e - \omega_p) \left[ \sigma_z , \sigma^- \right] + i g \left[ \sigma^+, \sigma^- \right] a
\end{align}
For a compact notation we will now use the notation $\Delta_{ab} = \omega_a - \omega_b$. With the commutators $\left[ \sigma_z, \sigma^-\right] = -2\sigma^-$, $\left[ \sigma^+, \sigma^- \right] = \sigma_z$, $\left[a^\dagger a, a \right] = -a$ and $\left[a, a^\dagger \right] = 1$, and adding in dissipative terms, we can simplify

\begin{align}
    &\dot{a} = - \left( \frac{\kappa_\mathrm{tot}}{2} + i \Delta_{rp} \right)a  - ig \sigma^- + a_\mathrm{in, 1} \sqrt{\kappa_1} \\
    &\dot{\sigma}^- = - \left(\Gamma_2 + i \Delta_{ep} \right) \sigma^- + i g \sigma_z a 
\end{align}
Here $\Gamma_2 = \gamma_1/2 + \Gamma_\varphi$ is the electron decoherence rate. For transmission spectra, we are only interested in the steady-state values of $a$ and $\sigma^-$, so we may set $\dot{a} = 0$ and $\dot{\sigma}^- = 0$. Furthermore, we will neglect resonator-qubit correlations by approximating $\langle \sigma_z  a \rangle \approx \langle \sigma_z \rangle \langle a \rangle$, valid for the weak probe conditions \cite{Blais2021}. Under these approximations we have

\begin{align}
    &-\left( \frac{\kappa_\mathrm{tot}}{2} + i \Delta_{rp} \right) \langle a \rangle  - ig \langle \sigma^- \rangle + a_\mathrm{in, 1} \sqrt{\kappa_1} = 0 \\
    &- \left(\Gamma_2 + i \Delta_{ep} \right) \langle\sigma^-\rangle + i g \langle \sigma_z \rangle \langle a \rangle = 0 \label{eq:neglect_correlator_a_sigmaz}
\end{align}

Solving this set of equations for $\langle a \rangle$ and $\langle \sigma^- \rangle$ gives
\begin{align}
    &\langle \sigma^- \rangle = \frac{g \langle \sigma_z \rangle}{\Delta_{ep} - i \Gamma_2} \langle a \rangle \\
    &\langle a \rangle  = \frac{a_\mathrm{in, 1} \sqrt{\kappa_1}}{\kappa_\mathrm{tot} / 2 + i \Delta_{rp} + i \chi(\omega_e) \langle \sigma_z \rangle} \label{eq:intracavity_field}
\end{align}
Eq.~\eqref{eq:intracavity_field} describes the intracavity field including coupling to the electron via the susceptibility $\chi (\omega_e)$:
\begin{align}
    \chi(\omega_e) = \frac{g^2}{\Delta_{ep} - i\Gamma_2}.
    \label{eq:susceptibility}
\end{align}
To translate this into a transmission spectrum we use the standard input-output theory of a two-sided cavity (see figure~\ref{fig:supp_input_output_model}):
\begin{align}
    a_\mathrm{in, 1} + a_\mathrm{out, 1} = \sqrt{\kappa_1} \langle a \rangle \\
    a_\mathrm{in, 2} + a_\mathrm{out, 2} = \sqrt{\kappa_2} \langle a \rangle
\end{align}
Since experimentally, we do not apply a probe tone at port 2 ($a_\mathrm{in, 2} = 0$) these equations simplify to 
\begin{align}
    a_\mathrm{out, 1} &= \sqrt{\kappa_1} \langle a \rangle - a_\mathrm{in, 1} \\
    a_\mathrm{out, 2} &= \sqrt{\kappa_2} \langle a \rangle. \label{eq:a_out_2_no_crosstalk}
\end{align}
These equations combined with Eq.~\eqref{eq:intracavity_field} can be used to describe the reflection and transmission spectrum:
\begin{align}
    S_{11} &= -1 + \frac{\kappa_1}{\kappa_\mathrm{tot} / 2 + i \Delta_{rp} + i \chi \langle \sigma_z \rangle} 
\end{align}
\begin{align}
    S_{21} &= \frac{a_\mathrm{out, 2}}{a_\mathrm{in, 1}} = \frac{\sqrt{\kappa_1 \kappa_2}}{\kappa_\mathrm{tot} / 2 + i \Delta_{rp} + i \chi \langle \sigma_z \rangle}.
    \label{eq:complexS21}
\end{align}
These results are consistent with Refs.~\cite{Cottet2017, Bonsen2023}. 

\subsection*{Origin of the asymmetric transmission spectrum}
\begin{figure*}
    \centering
    \includegraphics[width=0.6\textwidth]{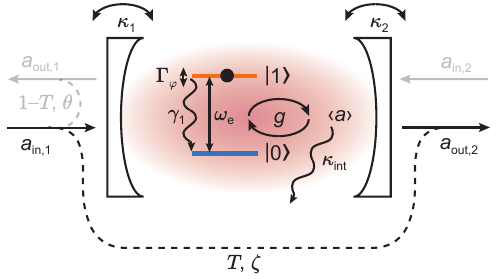}
    \caption{\textbf{Modeling experimental transmission spectra} Schematic displaying the resonator with input and output coupling rates $\kappa_{1,2}$, internal loss rate $\kappa_\mathrm{int}$ and intra-cavity field $\langle a \rangle$. The electron is represented as a two-level system with dephasing $\Gamma_\varphi$ and transverse decay $\gamma_1$, which couples to the resonator with rate $g$. The resonator is probed with strength $a_\mathrm{in, 1}$ and transmission is measured via $a_\mathrm{out, 2}$. In this experiment, we do not apply an input to $a_\mathrm{in,2}$ and do not detect at $a_\mathrm{out,1}$, so we may neglect these ports in our analysis. We include additional transmission crosstalk with amplitude $T$ and phase $\zeta$, and reflection crosstalk with amplitude $1-T$ and phase $\theta$.}
    \label{fig:supp_input_output_model}
\end{figure*}

The experimentally observed transmission spectrum is asymmetric (see figure~\ref{fig:S21_fit_with_parasitic}), which generally originates from impedance mismatches~\cite{Khalil2012}, coupling to nearby electromagnetic modes~\cite{Iizawa2021} or direct input-output cross-talk~\cite{Rieger2023}. Regardless of the exact origin of the asymmetry, we follow the theoretical description of Ref.~\cite{ViennotThesis2014} to incorporate crosstalk that bypasses the resonator, see figure~\ref{fig:supp_input_output_model}. This background crosstalk is given by a scattering matrix 
\begin{align}
    S = e^{i \zeta}  \begin{pmatrix} \sqrt{1 - T} e^{i \theta} & -i \sqrt{T} \\ -i \sqrt{T} & \sqrt{1 - T} e^{-i \theta}  \end{pmatrix},
\end{align}
where the diagonal elements are reflections from each port, and off-diagonal elements are the transmission crosstalk. We now modify $S_{21}$ by adding the transmission crosstalk to Eq.~\eqref{eq:a_out_2_no_crosstalk}:
\begin{align}
    a_\mathrm{out, 2} = -i\sqrt{T} e^{i\zeta} a_\mathrm{in, 1} + \sqrt{\kappa_2} \langle a \rangle.
\end{align}
This results in an additional frequency-independent term in $S_{21}$,
\begin{align}
     S_{21} &= S_{21}^\mathrm{parasitic} + S_{21} ^ \mathrm{res} \notag \\
     &= -i\sqrt{T} e^{i\zeta} + \frac{\sqrt{\kappa_1 \kappa_2}}{\kappa_\mathrm{tot} / 2 + i\Delta_{rp} + i \chi \langle \sigma_z \rangle}, \label{eq:s21_with_parasitic_transmission}
\end{align}
and it is the interference of these two terms that explains the asymmetric transmission spectrum~\cite{Rieger2023}.

\subsection*{Correcting for interference from background microwave transmission}
\begin{figure*}
    \centering
    \includegraphics[width=\textwidth]{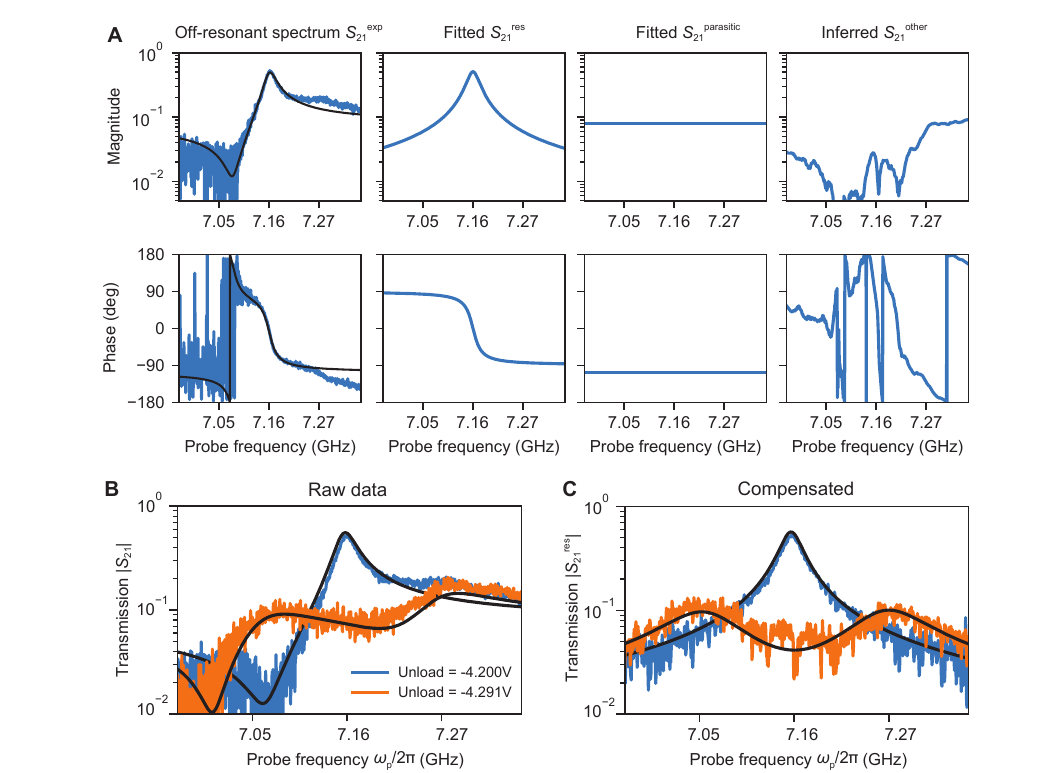}
    \caption{\textbf{Method for compensating for background transmission} (\textbf{A}) Far detuned resonator transmission spectrum (blue) with fit to Eq.~\eqref{eq:s21_with_parasitic_transmission} (black), yielding $T = 0.008 \pm $, $\zeta = -0.30\pm0.02$~rad. From the fit we reconstruct $S_{21}^\mathrm{res}$ and $S_\mathrm{21}^\mathrm{parasitic}$ and infer $S_{21}^\mathrm{other}$ as described in Eq.~\eqref{eq:residual_parasitic_transmission}. (\textbf{B}) Raw resonator spectra for two voltages where $\omega_e \gg \omega_r$ (blue) and $\omega_e \approx \omega_r$ (orange). The solid black lines are predictions from Eq.~\eqref{eq:s21_with_parasitic_transmission} with values of $T$ and  $\zeta$ from A. (\textbf{C}) Resonator transmission spectra from B, compensated for the parasitic transmissions, isolating the resonator-electron subsystem. Solid black lines are predictions from Eq.~\eqref{eq:s21_with_parasitic_transmission} with values $T= \zeta = 0$.}
    \label{fig:S21_fit_with_parasitic}
\end{figure*}

While Eq.~\eqref{eq:s21_with_parasitic_transmission} captures the essence of the transmission spectrum, there is a substantial deviation between fit and experimental data at higher probe frequencies (see figure~\ref{fig:S21_fit_with_parasitic}\,A), which complicates fitting to electron parameters $g$ and $\Gamma_2$ when $\omega_e \approx \omega_r$. This suggests that the experimental data consists of three components: $S_{21}^\mathrm{res}$, $S_{21}^\mathrm{parasitic}$ and additional $S_{21}^\mathrm{other}$, where the latter is due to additional transmission not captured by $T$ and $\zeta$. Crucially, $S_{21}^\mathrm{other}$ does not depend on voltage (see Fig.~2). To facilitate fitting of nearly resonant transmission spectra, we present a method to eliminate these additional transmission effects from our experimental data is similar to the background subtraction method of Ref.~\cite{YangPing2020}:

\begin{enumerate}
    \item For a far detuned ($\omega_e \gg \omega_r$) experimental transmission spectrum $S_{21}^\mathrm{exp}$, fit to Eq.~\eqref{eq:s21_with_parasitic_transmission} to determine $S_{21}^\mathrm{res}$ and $S_{21}^\mathrm{parasitic}$ by fitting only to the data near the peak frequency.
    \item Determine the additional parasitic component $S_{21}^\mathrm{other}$ from the fit residuals
    \begin{align}
        S_{21}^\mathrm{other} = S_{21}^\mathrm{exp} - S_{21}^\mathrm{res} - S_{21}^\mathrm{parasitic} \label{eq:residual_parasitic_transmission}
    \end{align}
    \item Because we assume $S_{21}^\mathrm{other}$ and $S_{21}^\mathrm{parasitic}$ do not depend on voltage, we may subtract $S_{21}^\mathrm{parasitic} + S_{21}^\mathrm{other}$ from measured transmission spectra at arbitrary voltage, which leaves only the information of the resonator-electron subsystem:
    \begin{align}
        S_{21}^\mathrm{res} = S_{21} ^ \mathrm{exp} - S_{21}^\mathrm{parasitic} - S_{21}^\mathrm{other}.
    \end{align}
\end{enumerate}
Figure~\ref{fig:S21_fit_with_parasitic}\,C demonstrates this compensation method for an off-resonant spectrum ($\omega_e  \gg \omega_r$, blue) and nearly resonant spectrum ($\omega_e \approx \omega_r$, orange). The nearly resonant spectrum shows the characteristic double peak vacuum Rabi splitting with peaks of approximately equal height, suggesting that the parasitic transmission was effectively removed and is indeed voltage independent. The raw data (figure~\ref{fig:S21_fit_with_parasitic}\,B) and compensated data both fit well to a model with parameters $g/2\pi~=~118$~MHz and $\Gamma_2 / 2\pi = 75$~MHz, which further suggests that the compensation method does not introduce a bias when determining $g$ and $\Gamma_2$.

\section{Estimating electron motional state frequencies from resonator transmission spectra}
\label{supp:estimating_fe_from_resonator_transmission}
Besides direct measurement with two-tone spectroscopy, the electron motional state frequency can also be estimated from the measured resonator frequency $\omega_{r}$. Here, we explain this method, which is used for Fig.~\ref{fig:fig3}\,B in the main text. 

\begin{figure*}
    \centering
    \includegraphics[width=0.5\textwidth]{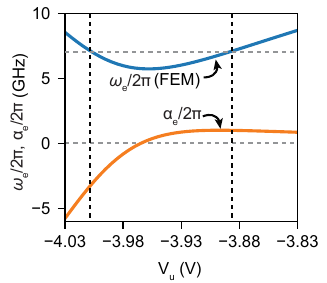}
    \caption{\textbf{Anharmonicity of the trapping potential of the quantum dot} Anharmonicity $\alpha_{e}= \omega_{12}-\omega_{01}$ and simulated electron frequency $\omega_{e}$ of the compensated electrostatic trapping potential from Fig.~\ref{fig:fig3}\,B. Values are calculated from FEM using the values from table~\ref{tab:compensating_voltages_1d}. Horizontal dashed line indicate where $\alpha_{e} = 0$ and $\omega_{e} = \omega_{r}$, and vertical dashed lines indicate the voltages where $\omega_{e} \approx \omega_{r}$. The anharmonicity is large at most voltages, supporting the use of a two-state model for the electron used to produce the fit from the spectroscopy data shown in Fig.~\ref{fig:fig3}\,B. The anharmonicity only vanishes near the minima of the electron frequency, which may explain some of the broadening in the two-tone signal seen in Fig.~\ref{fig:fig3}\,B.} 
    \label{fig:supp_anharmonicity_two_tone}
\end{figure*}

The resonator frequency shift in the off-resonant regime is given by the real part of the electron susceptibility~\cite{Cottet2017}, as defined in Eq.~\eqref{eq:susceptibility}: 
\begin{align}
    \Delta \omega_r = \mathrm{Re}(\chi \langle \sigma_z \rangle) = \frac{g^2 \Delta_{er} \langle \sigma_z \rangle}{\Delta_{er}^2 + \Gamma_2^2}.
\end{align}
When the electron-resonator detuning $\Delta_{er}$ is large relative to the electron linewidth, and if the electron remains in its ground state, the measured resonator frequency shift $\Delta \omega_{r}$ is approximated by $\Delta \omega_{r} = g^{2}/\Delta_{er}$. Thus, the inferred electron frequency is given by 
\begin{align}
    \omega_{e}=\omega_{r}+g^{2}/\Delta \omega_{r},
    \label{eq:omega_e_estimation_2_levels}
\end{align}
where $\omega_{r}/2\pi$ is the bare resonator frequency uncoupled from the electron. The electron frequency calculated from the resonator spectroscopy measurement is displayed in Fig.~\ref{fig:fig3}\,B. For this calculation, we find the best agreement between the two-tone spectroscopy data and the extracted electron frequency from resonator spectroscopy for $g/2\pi=110$MHz. The bare resonator frequency $\omega_r$ is determined with an empty dot and the model further assumes that only two electron states couple to the resonator. For most voltage configurations, the anharmonicity of the system is high (figure~\ref{fig:supp_anharmonicity_two_tone}) and justifies this use of a two-state model for the electron used to produce the fit from the spectroscopy data in Fig.~\ref{fig:fig3}\,B.

In Fig.~\ref{fig:fig3}\,B we compare the estimate using Eq.~\eqref{eq:omega_e_estimation_2_levels} with two-tone spectroscopy, which show good agreement. This lends further credibility to our use of a two-level approximation method for Fig.~\ref{fig:fig3}\,B. 



\section{Analytical model of electron motional state spectrum}
\label{supp:analytical_model_of_electron_spectrum}
The previous section described how to accurately model the measured electron frequency curve shown in Fig.~\ref{fig:fig3}\,B using FEM. To gain further understanding about the experimental electrostatic potential and the trapped electron, we study a simple model with three ingredients: (a) a harmonic confinement (b) a constant electric field (c) an anharmonic term. We show that the combination of these three ingredients qualitatively modifies the voltage dependence from $\omega_{e} \propto \sqrt{V}$ to what is observed in the experiment. It also reproduces the minimum frequency observed in the experiment.

The simple one dimensional electrostatic potential that reproduces the experimental data is given by
\begin{align}
    U = -eV(y)=  a_1 y^2 + a_2 y^4 - e E_y y,\label{eq:ex_perturbation_1}
\end{align}
where $E_y$ is the in-plane electric field, $a_1$ is the harmonic term, and $a_2$ contains the anharmonicity. From the FEM model we have observed that when all other electrodes are biased appropriately, the Unload electrode sweeps the harmonic component $a_1 \propto V_{u}$ through 0. To calculate the effect of all terms on the effective motional frequency, we differentiate $U$ twice with respect to $y$:
\begin{align}
    \tilde{\omega}_e^2 = \frac{e}{m_e} \frac{\partial^2 U}{\partial y^2} = \frac{2a_1}{m_e} + \frac{12 a_2}{m_e} y_0^2,  \label{eq:omega_with_in_plane_field}
\end{align}
where $y_0$ is the coordinate of the potential minimum. 

In the absence of anharmonicity ($a_2 = 0$), we retrieve $\tilde{\omega}_e^2 = 2a_1/m_e$ and since $a_1$ varies linearly with the applied electrode voltage, $\tilde{\omega}_e \propto \sqrt{V}$ as shown by the blue curve in figure~\ref{fig:supp_qualitative_model}\,A. 

If the anharmonicity is non-zero, we must calculate $y_0$ by differentiating the electrostatic potential \eqref{eq:ex_perturbation_1} once with respect to $y$:
\begin{align}
    \frac{\partial U}{\partial y}\bigg\rvert_{y=y_0}  = 2 a_1 y_0 + 4 a_2 y_0^3 - e E_y = 0. \label{eq:in_plane_x0_condition}
\end{align}
The resulting equation for $y_0$ is 
\begin{align}
     y_0^3 + \frac{a_1}{2 a_2} y_0 - \frac{e E_y}{4 a_2} = 0, \label{eq:x0_condition}
\end{align}
which is of the form $x^3 + p x + q = 0$, and is also known as a depressed cubic equation. From inspection, we see that
\begin{align}
    p &= \frac{a_1}{2 a_2} \\ 
    q &= -\frac{e E_y}{4a_2} \\
    \frac{p}{q} &= -\frac{2a_1}{e E_y}.
\end{align}

The exact solution to this equation can be classified according to the sign of the determinant $D$, which is defined as
\begin{align}
    D &= -(4p^3 + 27 q^2)
\end{align}

\begin{itemize}
    \item If $D > 0$ the cubic has three distinct real solutions. This happens if the potential has a clear double well shape (the third solution is the bump in between the wells), for example. In practice, this happens for negative anharmonicity $a_2 < 0$ and for small $E_y$.
    \item If $D < 0$, the cubic has one real solution and two complex conjugate pair solutions. This happens if the potential has one clear minimum, either because $a_2 > 0$ or because $E_y$ is large (in this case the sign of $a_2$ is irrelevant).
\end{itemize}

For now let us focus on the second case, where there is one \emph{real} solution. 

\subsection*{Exact analytical case}
The exact case is analytically tractable and Cardano's formula prescribes a solution. The potential minimum location is given by 
\begin{align}
    y_0 &= \left( -\frac{q}{2} + \sqrt{\frac{q^2}{4} + \frac{p^3}{27}} \right)^{\frac{1}{3}} + \left(-\frac{q}{2} - \sqrt{\frac{q^2}{4} + \frac{p^3}{27}} \right)^{\frac{1}{3}} \label{eq:cardanos_formula_depressed_qubic}\\
    &= \left(\frac{e E_y}{8a_2} + \sqrt{\frac{- D}{108}}\right)^{\frac{1}{3}}  + \left( \frac{e E_y}{8a_2}  - \sqrt{\frac{- D}{108}} \right)^{\frac{1}{3}}
\end{align}
In evaluating this expression, we discard any complex expressions from taking the cube root, since we are interested in real values of $y_0$ only. 

\subsection*{Estimate of the minimum electron frequency}
To estimate the minimum electron frequency predicted by this model, we expand the square root in Eq.~\eqref{eq:cardanos_formula_depressed_qubic} to first order in the small quantity $p/q$, which is valid because near the minimum of the electron frequency curve $p/q \propto a_1/E_y \ll 1$. This yields: 
\begin{align}
    y_0 &\approx \left[ \frac{p^3}{27q} \right]^\frac{1}{3} + \left[ -q \left( 1 + \frac{p^3}{27 q^2} \right) \right]^\frac{1}{3} \\
    &\approx -q^\frac{1}{3} + \frac{q^\frac{2}{3}}{3} \left( \frac{p}{q} \right) - \frac{p}{3} \frac{q^\frac{1}{3}}{27} \left( \frac{p}{q} \right)^2 ,\label{eq:supp_y0_first_order}
\end{align}
which is consistent with the zero-th order solution $y_0 = -q^{1/3}$, obtained by setting $p =0$ in Eq.~\eqref{eq:x0_condition}. Plugging this value for $y_0$ into Eq.~\eqref{eq:omega_with_in_plane_field} gives an estimate for the minimum frequency due to in-plane fields
\begin{align}
    \tilde{\omega}_\mathrm{min}^2 \approx \frac{12 a_2}{m_e} \left( q^\frac{2}{3} - \frac{2p}{3} \right) \approx \frac{12a_2^\frac{1}{3}}{m_e} \left(\frac{e E_y}{4}  \right)^\frac{2}{3} ,
\end{align}
where in the last simplification we have used the definition of $q$. Note that if either $E_y =0$ or $a_2 =0$, $\omega_\mathrm{min} = 0$. 

\begin{figure*}
    \centering
    \includegraphics[width=\textwidth]{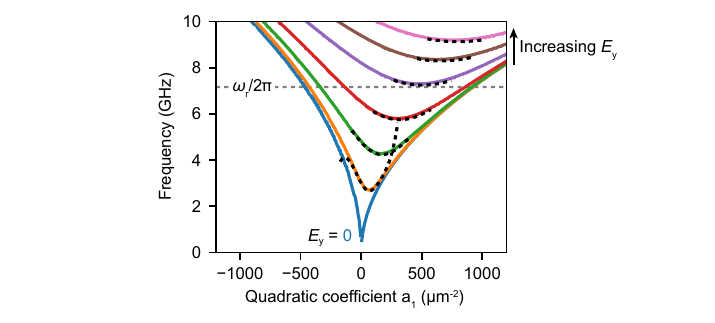}
    \caption{\textbf{Qualitative model for the measured electron spectrum}  Modeled electron frequency as function of harmonic coefficient $a_1$, for fixed $a_2$. Colored curves represent increasing values of $E_y$, which show a lifting of the minimum electron frequency from zero. Black dashed lines are first order approximations from Eq.~\eqref{eq:supp_y0_first_order}. This simple model reproduces the key experimental observations in Fig.~\ref{fig:fig3}\,B.} 
    \label{fig:supp_qualitative_model}
\end{figure*}

In figure~\ref{fig:supp_qualitative_model} we show that the minimum electron frequency increases for larger in-plane fields $E_y$, and for very large $E_y$ a crossing with the resonator becomes impossible. The experimentally observed minimum electron frequency of approximately 6~GHz corresponds to a moderate value of $E_y$. This model qualitatively reproduces the main features of the electron frequency curve observed in the experiment, namely the deviation of a square-root dependence and the observation of a nonzero minimum frequency. It also provides further evidence of an anharmonic trapping potential \emph{and} the presence of an in-plane field $E_y \neq 0$. 

\section{Purcell decay}
\label{supp:purcell_decay}
\subsection*{Decay through the resonator}
The Purcell decay through the resonator depends on the coupling strength $g$, resonator linewidth $\kappa$ and electron-resonator detuning $\Delta_{er} = \omega_e - \omega_r$ \cite{Schuster2010, Blais2004, Krantz2019}:
\begin{align}
    \gamma_1 = \frac{1}{T_1} = \frac{g^2 \kappa}{\Delta_{er}^2 + (\kappa / 2)^2} \label{eq:T1_contrib_resonator}
\end{align}
There are two limiting cases of interest. First, on resonance ($\omega_e = \omega_r$) $\gamma_1 = 4 g^2 / \kappa$ resulting in a sharp drop in $T_1$. Secondly if $\Delta_{er} \gg \kappa$, $\gamma_1 = (g/\Delta)^2 \kappa$. 

\begin{figure*}
    \centering
    \includegraphics[width=0.4\textwidth]{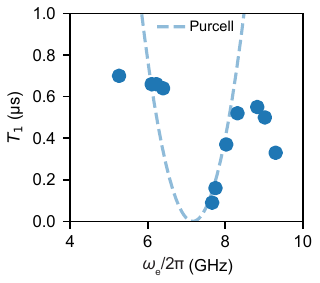}
    \caption{\textbf{$T_{1}$ dependence on electron frequency} Measured $T_{1}$ times at various electron frequencies. In these measurements the electron frequency is tuned by varying $V_{u}$. The dashed line is the expected Purcell decay rate through the resonator calculated from Eq.~\eqref{eq:T1_contrib_resonator}. Good agreement for $\Delta_{er}<$ 1.1 GHz suggests that $T_{1}$ is limited by Purcell decay for these values of the detuning, but that other mechanisms limit $T_{1}$ for $\Delta_{er}>$ 1.1 GHz.} 
    \label{fig:supp_T1_voltage_purcell}
\end{figure*}

We numerically evaluate Eq.~\eqref{eq:T1_contrib_resonator} in figure~\ref{fig:supp_T1_voltage_purcell} for $g / 2\pi = 110$~MHz, $\kappa / 2\pi = 23$~MHz and $\omega_r / 2\pi = 7.163$~GHz. The predicted $T_1$ values drop below 1~$\mu$s for $\Delta_{er} < 1.1$~GHz and the agreement between experiment and theory suggests that $T_1$ is limited by Purcell decay for such small $\Delta_{er}$. However, the discrepancy for $\Delta_{er} > 1.1$~GHz suggests there are other mechanisms limiting the measured $T_1$ for large values of detuning.

\subsection*{Decay through other bias electrodes}
\begin{figure*}
    \centering
    \includegraphics[width=0.6\textwidth]{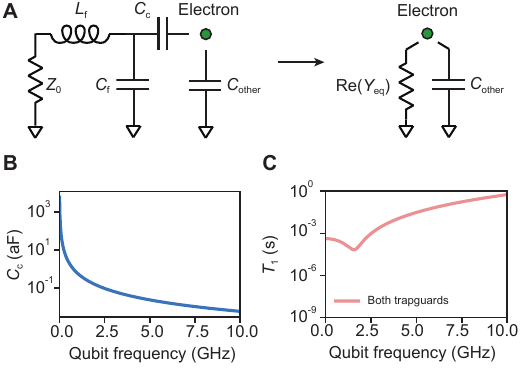}
    \caption{\textbf{Purcell decay through the bias electrodes} (\textbf{A}) Circuit model for calculating the $T_1$ contribution due to bias electrodes. The trapped electron is capacitively coupled to electrodes with an on-chip filter (inductance $L_f$ and capacitance $C_f$) and the load impedance of the voltage source is $Z_0$. The real part of the equivalent admittance acts as a resistor which affects $T_1$. (\textbf{B}) Calculated coupling capacitance for the Guard electrodes  (\textbf{C}) Corresponding $T_1$ contribution for the Guard electrodes.}
    \label{fig:T1_bias_electrodes}
\end{figure*}

We calculate the $T_1$ contribution due to a single dot electrode from the circuit model in figure~\ref{fig:T1_bias_electrodes}\,A. This model assumes the electron couples capacitively to a bias electrode with capacitance $C_c$, and has additional capacitance to ground, or other electrodes specified by $C_\mathrm{other}$. As in the experiment, the bias electrode also contains an on-chip filter and is connected to a load impedance $Z_0 = 50$~$\Omega$. Similar to the calculation of a Purcell effect in superconducting qubits~\cite{Houck2008}, we calculate the decay rate through the bias electrode via the effective bias electrode admittance $Y_\mathrm{eq}$ (see figure~\ref{fig:T1_bias_electrodes}\,A): 
\begin{align}
    \gamma_1 = \frac{1}{T_1} = \frac{\mathrm{Re}(Y_\mathrm{eq} (\omega_e))}{C_\mathrm{other}},    
\end{align}
where we neglect a renormalization of the capacitance $C_\mathrm{other}$ due to the imaginary part of $Y_\mathrm{eq}$. After expressing $Y_\mathrm{eq}$ in terms of circuit parameters, we find 
\begin{align}
    \gamma_1 = \omega_e \frac{C_c}{C_\mathrm{other}} \frac{\omega_e Z_0 C_c}{\left( 1 - \omega_e^2 L_f (C_f + C_c)\right)^2 + \left( \omega_e Z_0 (C_f + C_c) \right)^2},    
\end{align}
where $0 \leq C_c / C_\mathrm{other} \leq 1$. It is clear that $\gamma_1$ depends sensitively on $C_c$, so it is important to determine the value of $C_c$ accurately.

We determine a reasonable value for the capacitance $C_c$ following an argument presented in Ref.~\cite{Castoria2025}. Suppose the electron is confined at the center of a one-dimensional harmonic potential with confinement energy $\frac{1}{2} m_e \omega_e^2 y^2$. A small voltage $\Delta V$ on the bias electrode produces an electric field $E$ at the location of the electron which changes its position by $\Delta y$. Newton's second law states 
\begin{align}
    m_e \omega_e^2 \Delta y = e E = e \Delta V  \frac{\partial \alpha}{\partial y}, \label{eq:newton_second_law}
\end{align}
where $\alpha$ is the coupling constant of the bias electrode at the electron's location. From this we find 
\begin{align}
    \Delta y = \frac{e \Delta V}{m_e \omega_e^2} \frac{\partial \alpha}{\partial y}.
\end{align}
The induced charge due to the change in electron position is given by
\begin{align}
    \Delta q &= e \frac{\partial \alpha}{\partial y} \Delta y = \frac{e^2}{m_e \omega_e^2} \left( \frac{\partial \alpha}{\partial y} \right)^2 \Delta V.
\end{align}
The capacitance of an electron to a bias electrode is therefore
\begin{align}
    C_c = \frac{\Delta q}{\Delta V} = \frac{e^2}{m_e \omega_e^2} \left( \frac{\partial \alpha}{\partial y}\right)^2. \label{eq:C_e_for_bias_electrodes}
\end{align}
In principle, Eq.~\eqref{eq:C_e_for_bias_electrodes} applies to arbitrary bias electrodes (including the microwave resonator), however Eq.~\eqref{eq:C_e_for_bias_electrodes} is only valid for resonator bias electrode if $\omega_e \gg \omega_r$, because the coupling to the resonator modifies Newton's second law (Eq.~\eqref{eq:newton_second_law}) \cite{Castoria2025}. 

From finite element modeling we find that the Guard electrodes have the dominant capacitance with $\partial \alpha / \partial y \approx 0.03\, \mu$m$^{-1}$, which is small compared with $\partial \alpha / \partial y \approx 0.46\, \mu$m$^{-1}$ for each resonator electrode. The Bottom and Unload electrodes are even less sensitive with $\partial \alpha / \partial y \approx 0$, suggesting the Guards are the only relevant bias electrodes to consider. With a filter inductance of $L_f = 12$~nH and capacitance of $C_f = 0.8$~pF, the resulting electron capacitance and $T_1$ contribution (conservatively assuming $C_c / C_\mathrm{other} = 1$) are shown in figure~\ref{fig:T1_bias_electrodes}\,B,C. The predicted $T_1 > 1$~ms for $\omega_e / 2\pi > 3$~GHz, with a slight dip in predicted $T_1$ due to a filter resonance at $\omega = 1/\sqrt{L_f C_f} \approx 2\pi \times2$~GHz.

\section{Classical helium vibrations}
\label{supp:classical_helium_vibrations}
\begin{figure*}
    \centering
    \includegraphics[width=0.9\textwidth]{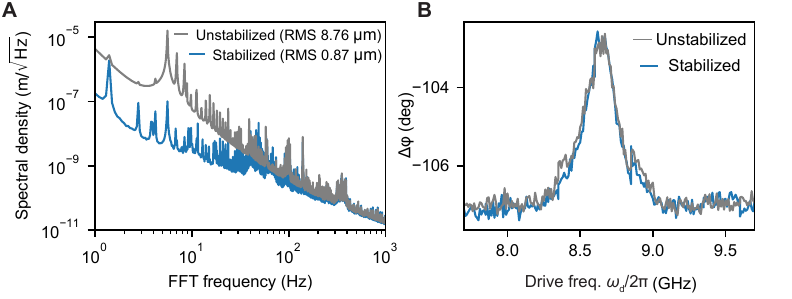}
    \caption{\textbf{Vibrational noise spectra and electron state two-tone linewidth with and without cryostat vibration stabilization} (\textbf{A}) Vibrational noise spectra acquired using a geophone sensor when cryostat active vibration cancellation system is on and off. Peaks in both spectra occur at harmonics of the pulse tube at 1.4 Hz. RMS noise amplitude undergoes a ten-fold decrease from 8.76 $\mu$m to 0.87 $\mu$m when stabilization is turned on. (\textbf{B}) Despite the ten-fold decrease in the vibration noise spectral density when the active stabilization is turned on, the measured two-tone linewidth of the electron motional state $\gamma$ does not change.}
    \label{fig:supp_vibration_spectra}
\end{figure*}

Figure~\ref{fig:supp_vibration_spectra}\,A shows the noise spectral density of vibrations of our cryostat acquired using a geophone when the active stabilization is on and off. Although the low-frequency noise spectral density increases by an order of magnitude when the active stabilization is turned off, we observe no change in the measured two-tone linewidth (figure~\ref{fig:supp_vibration_spectra}\,B).

\section{Temperature-dependent observation of vacuum Rabi splitting}
\label{supp:temp_dependent_rabi_splitting}

\begin{figure*}
    \centering
    \includegraphics[width=\textwidth]{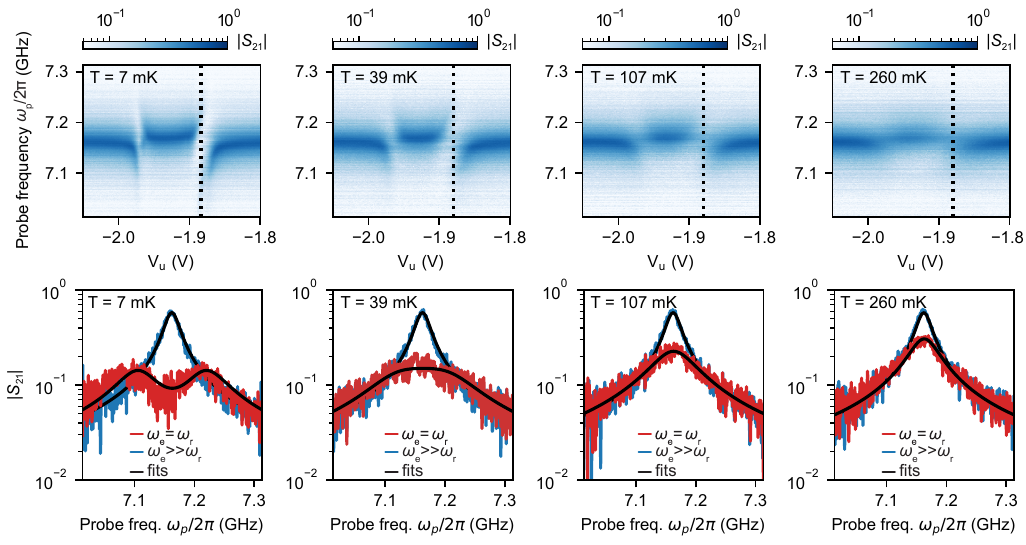}
    \caption{\textbf{Emergence of the vacuum Rabi splitting with decreasing temperature} (Top)~Resonator spectroscopy measurements taken at $T=7, 39, 107, \textrm{and}~260$~mK. The prominence of the anticrossings decreases as the system gets warmer. Dashed vertical line is where the electron frequency is on resonance with the resonator. (Bottom) Resonator transmission spectrum taken off-resonance (blue) and on-resonance at the dashed black line (red) for the spectroscopy measurement taken at each temperature. The Rabi splitting is only clearly visible at the low temperatures. The spectra are compensated for parasitic transmission and fit to the transmission model as described in the Supplementary Text, S\ref{supp:modeling_resonator_transmission_spectra}.}
    \label{fig:supp:splitting_vs_temp}
\end{figure*}

\begin{figure*}
    \centering
    \includegraphics[width=\textwidth]{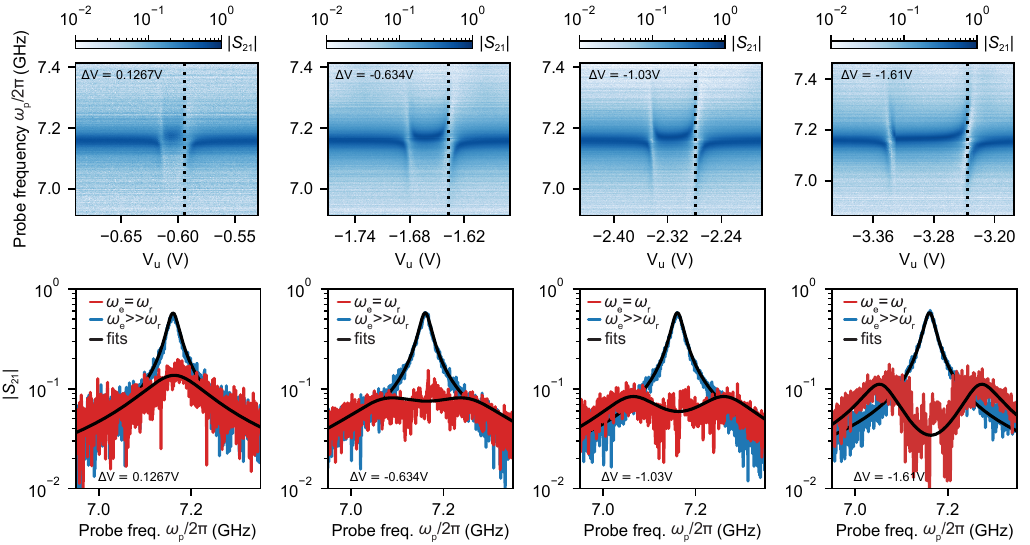}
    \caption{\textbf{Emergence of the vacuum Rabi splitting with increasing voltage offset} (Top)~Resonator spectroscopy measurements taken at $\Delta V=$ 0.1267, -0.634, -1.03, -1.61V. The prominence of the anti-crossings increases as the voltage offset becomes more negative. Dashed vertical line is where the electron frequency is on resonance with the resonator. (Bottom) Resonator transmission spectrum taken off-resonance (blue) and on-resonance at the dashed black line (red) for the spectroscopy measurement taken at each temperature. The Rabi splitting is only clearly visible at more negative offsets. The spectra are compensated for parasitic transmission and fit to the transmission model as described in the Supplementary Text, S\ref{supp:modeling_resonator_transmission_spectra}.}
    \label{fig:supp:splitting_vs_offset}
\end{figure*}

We see experimentally that the prominence of the vacuum Rabi splitting we observe in resonator spectroscopy measurements improves as both the temperature decreases (figure~\ref{fig:supp:splitting_vs_temp}), and as the negative offset applied to the dot electrodes increases (figure~\ref{fig:supp:splitting_vs_offset}). The temperature-dependent manifestation of this vacuum Rabi splitting is consistent with coupling of the trapped electron to a thermally activated bath, which could originate from thermal fluctuations of the helium, or from a collection of additional charges in the vicinity of the dot.

\section{Prospects for spin readout}
\label{supp:prospects_for_spin_coupling}
It is natural to ask what the results in this manuscript imply for coupling to the electron on helium spin state~\cite{Lyon2006}. One way to couple a single spin to a microwave resonator is by applying a magnetic field gradient to hybridize the charge motion and spin degrees of freedom in a manner analagous to  that which has been achieved previously in semiconducting quantum dots~\cite{Samkharadze2018,Mi2018}. In addition, this approach has been studied theoretically for both semiconducting quantum dots, as well as for electrons on helium~\cite{Schuster2010, Dykman2023} and neon~\cite{Tian2025} and represents a promising path toward achieving electron spin readout. 

An experimental demonstration of spin-photon coupling involves sweeping the spin qubit frequency through the microwave resonator frequency while the charge qubit frequency is detuned off-resonance to isolate the spin state from charge noise. Below, we show this detuning dilutes the spin-photon coupling by a factor ($g_{cs}/\Delta_{cs} < 1$), but the suppression of charge noise scales as $(g_{cs}/\Delta_{cs})^2$~\cite{Burkard2020, Tian2025}, which implies a large $\Delta_{cs}$ is generally preferable.

The spin-photon coupling strength $g_s$ can be written as 
\begin{align}
    g_s &= g_c \frac{\mu_B a_x}{\hbar} \frac{\partial B_z}{\partial x} \frac{\sqrt{2}}{ \omega_c (1 - \omega_s^2 / \omega_c^2)} \\
    &= g_c \frac{\mu_B a_x}{\hbar} \frac{\partial B_z}{\partial x} \frac{\sqrt{2} \omega_c}{ (\omega_c - \omega_s)(\omega_c + \omega_s)} \\
    &\approx g_c \frac{\mu_B a_x}{\sqrt{2}\hbar} \frac{\partial B_z}{\partial x} \frac{1}{\omega_c - \omega_s} \\
    &= g_c \, \frac{g_{cs}}{\Delta_{cs}} \label{eq:spin_photon_hybridization}, 
\end{align}
where $\mu_B$ is the Bohr magneton, $g_c/2\pi$ is the charge-photon coupling strength, $a_x = \left(\hbar / m_e \omega_c \right)^\frac{1}{2} \approx 50$~nm is the spatial extent of the charge ground state wavefunction, $\omega_c/2\pi$ is the charge qubit frequency, $\omega_s/2\pi$ is the spin qubit frequency set by the Zeeman splitting, and $\partial B_z/\partial x$ is the magnetic field gradient. 

Eq.~\eqref{eq:spin_photon_hybridization} shows that the spin-photon coupling is proportional to the charge-photon coupling, and is reduced by a numerical factor $g_{cs} / \Delta_{cs}$, where 
\begin{align}
    \frac{g_{cs}}{2\pi} = \frac{\mu_B a_x}{\sqrt{2} h} \frac{\partial B_z}{\partial x} \label{eq:spin_charge_hybridization_strength}
\end{align}
Note that Eq.~\eqref{eq:spin_charge_hybridization_strength} can also be derived from the theory developed for semiconducting double quantum dots~\cite{Benito2017, Tian2025}, by setting $2t_c/h = \omega_c$ and taking the approximation $g_c/\Delta_{cs} \ll 1$. Using $\partial B_z/\partial x = $~0.1 mT/nm for the field gradient (see e.g. Refs.~\cite{Schuster2010}~and~\cite{Tian2025}), the spin-charge coupling evaluates to $g_{cs}/2\pi \approx 50$~MHz. From the value of $g_c/2\pi \simeq 120$~MHz measured in this work, we infer that the charge qubit can be detuned by $\Delta_{cs}/2\pi\approx 2$~GHz while still achieving spin coupling rate of $g_s/2\pi \approx 3$~MHz. As such, microsecond spin-qubit readout should be within reach.

\clearpage
\newpage
\clearpage


\begin{table*} 
    \centering
    \caption{\textbf{Voltage and E-field parameters used in FEM simulation of electron frequency.} ``Unadjusted'' refers to experimentally used values, ``Voltage only'' refers to adjustments in FEM with offsets in dot electrode voltages only, ``$E_{y}$ only'' refers to adjustments in FEM with compensating in-plane electric field only, ``Voltage + $E_{y}$'' refers to adjustments in both voltage offset and electric field used to produce ``Simulated'' best fit in Fig.~\ref{fig:fig3}\,B. \\} 
    \label{tab:compensating_voltages_1d} 

    \begin{tabular}{|l|c|c|c|c|} 
        \hline
		FEM params. & Unadjusted & Voltage only & $E_{y}$ only & Voltage + $E_{y}$ \\
		\hline
		$V_{b}$ (V) & -1.68 & -1.58 & -1.68& -1.58\\
		$V_{g, up}$ (V) & -3.35 & -3.35 & -3.35 & -3.35\\
		$V_{g, down}$ (V) & -2.85 & -2.85 & -2.85 & -2.85\\
            $V_{r}$ (V) & -2.013 &-2.013  & -2.013  & -2.013 \\
            $V_{u}$ (V) & (-4.03, -3.83) & (-4.0, -3.8) & (-4.03, -3.83) & (-4.0, -3.8)\\
            $E_{y}$ (V/cm) & 0 & 0 & 142 & 142\\
            $E_{x}$ (V/cm) & 0 & 0 & 0 & 0\\
		\hline
    \end{tabular}
\end{table*}

\begin{table*} 
    \centering
    \caption{\textbf{Voltage and E-field adjustments used in FEM simulation of electron frequency.} ``Experimental" column values are those used in Fig.~\ref{fig:fig2}B. ``Adjusted'' column values are parameters used to produce dashed line resonance condition in figure~\ref{fig:supp_voltage_comp}\,B \\} 
    \label{tab:compensating_voltages_2d} 

    \begin{tabular}{|l|c|c|c|c|} 
        \hline
		FEM params. & Experimental & Adjusted \\
		\hline
		$V_{b} (V)$ & -1.67 & -1.555 \\
		$V_{g, up}$ (V) & -3.35 & -3.35\\
		$V_{g, down}$ (V) & -2.85 & -2.85 \\
            $V_{r}$ (V) & (-2.055, -2.015) & (-2.055, -2.015)\\
            $V_{u}$ (V) & (-4.5, -4.2) & (-4.55, -4.25) \\
            $E_{y}$ (V/cm) & -- & 172.9 \\
            $E_{x}$ (V/cm) & -- & -105.0 \\
		\hline
    \end{tabular}
\end{table*}

\begin{table*} 
    \centering
    \caption{\textbf{Experimental voltage configurations used when applying global offset} Voltages used in resonator spectroscopy sweeps from which $\Gamma_{2}$ values are extracted and presented in Fig.~\ref{fig:fig4}\,C. The $\Delta V$ parameter presented as the x-axis of Fig.~\ref{fig:fig4}\,C is defined as $V_{r}$.\\} 
    \label{tab:lifted_dot_voltages} 

    \begin{tabular}{|l|c|c|c|c|} 
        \hline
		$\Delta V$ (V) &$V_{b}$ (V) & $V_{r}$ (V) & ($V_{g,up}$, $V_{g,down}$) (V)  & $V_{u} (V)$\\
		\hline
		0.1267 &0.4  & 0.1267 & (-0.6, -0.6) & -0.594 \\
        -0.253 & 0  & -0.253 & (-1.0, -1.035) & -1.104 \\
        -0.634 & -0.4 & -0.634 & (-1.4, -1.44) & -1.637 \\
        -1.03 & -0.8 & -1.03 & (-1.8, -1.87) & -2.278 \\
        -1.415 & -1.2 & -1.415 & (-2.2, -2.275) & -2.887 \\
        -1.61 &-1.4 & -1.61 & (-2.4, -2.477) & -3.236 \\
		\hline
    \end{tabular}
\end{table*}

\end{document}